\documentclass[twocolumn,showpacs,preprintnumbers,amsmath,amssymb,prb,floatfix]{revtex4}
\usepackage{graphicx}
\usepackage{dcolumn}
\usepackage{bm}
\usepackage{amsmath,amsthm,amssymb,ascmac}
\usepackage{color}

\pacs{ 05.60.Gg,  72.25.-b, 73.63.Kv, 05.70.Ln}

\newcommand{\be}{\beta}
\newcommand{\bea}{\begin{eqnarray}}
\newcommand{\eea}{\end{eqnarray}}
\newcommand{\aeq}{\!\! &=& \!\!}
\newcommand{\aeqe}{\!\! & \equiv & \!\!}
\newcommand{\aeqd}{\!\! & \stackrel{\mathrm{def}}{=} & \!\!}

\newcommand{\aeqap}{\!\! &\approx & \!\!}
\newcommand{\bra}{\langle}
\newcommand{\dbra}{\langle  \hspace{-0.5 mm} \langle}
\newcommand{\ket}{\rangle}
\newcommand{\dket}{\rangle \hspace{-0.5 mm} \rangle }
\newcommand{\mR}{\mathcal{R}}

\newcommand{\mO}{\mathcal{O}}

\newcommand{\tot}{{\rm{tot}}}
\newcommand{\tr}{\mbox{Tr}}
\newcommand{\dg}{^\dagger}
\newcommand{\tl}{\tilde}

\newcommand{\defe}{\stackrel{\mathrm{def}}{=}}
\newcommand{\e}{\equiv}
\newcommand{\ep}{\varepsilon}
\newcommand{\ga}{\gamma}
\newcommand{\Ga}{\Gamma}
\newcommand{\ka}{\kappa}
\newcommand{\up}{\uparrow}
\newcommand{\dw}{\downarrow}
\newcommand{\al}{\alpha}
\newcommand{\bu}{\bullet}
\newcommand{\sig}{\sigma}

\newcommand{\f}{\frac}
\newcommand{\half}{\frac{1}{2}}
\newcommand{\pr}{\prime}
\newcommand{\dl}{\delta}
\newcommand{\Dl}{\Delta}
\newcommand{\lm}{\lambda}
\newcommand{\Lm}{\Lambda}

\newcommand{\om}{\omega}
\newcommand{\Om}{\Omega}
\newcommand{\sinc}{\mbox{sinc}}

\newcommand{\la}{\label}
\newcommand{\ke}[1]{ \vert #1 \rangle }
\newcommand{\br}[1]{ \langle #1 \vert }
\newcommand{\abs}[1]{\vert #1 \vert }
\newcommand{\no}{\nonumber}

\newcommand{\hc}{\mbox{h.c.}}

\newcommand{\re}[1]{Eq. (\ref{#1})}
\newcommand{\res}[1]{\S \ref{#1}}
\newcommand{\Ref}[1]{Ref. [\onlinecite{#1}]}
\newcommand{\hs}{\hspace}
\newcommand{\taiou}{ \ \ \ \longleftrightarrow  \ \ \  }

\newcommand{\dbr}[1]{ \langle  \hspace{-0.5 mm} \langle #1 \vert }
\newcommand{\dke}[1]{\vert #1 \rangle \hspace{-0.5 mm} \rangle }
\newcommand{\rd}[1]{\textcolor{black}{#1}}
\newcommand{\bl}[1]{\textcolor{black}{#1}}
\newcommand{\BL}[1]{\textcolor{black}{#1}}
\newcommand{\RD}[1]{\textcolor{black}{#1}}

\begin{document}

\title{Interaction effect on adiabatic pump of charge and spin in quantum dot}

\author{Satoshi Nakajima}
\author{Masahiko Taguchi}
\author{Toshihiro Kubo}
\author{Yasuhiro Tokura}
\email{tokura.yasuhiro.ft@u.tsukuba.ac.jp}

\date{\today}

\affiliation{
Graduate School of Pure and Applied Sciences, University of Tsukuba, 
1-1-1, Tennodai, Tsukuba, 305-8571, Japan}
\date{\today}

\begin{abstract}
We investigate the pumped charge and spin at zero bias by \RD{a modulation} of two control parameters using \rd{the full counting statistics with quantum master equation approach}. 
First we study \rd{higher order} effects of the pumping frequency in \rd{general Markov systems} and show \RD{in this limit} the equivalence between our approach and the real-time diagrammatic approach. 
\bl{An adiabatic modulation of the control parameters induces the Berry-Sinitsyn-Nemenman (BSN) phase. 
We show that the origin of the BSN phase is a nonadiabatic effect}. 
The \RD{pumped} charge (spin) is given by a summation of \bl{(i)} a time integral of the \rd{instantaneous steady} charge (spin) current and \bl{(ii)} a geometric surface integral of 
the \rd{BSN} curvature\bl{, which results from the BSN phase. In quantum dots (QDs) weakly coupled to two leads}, 
we show that \bl{(i)} \rd{is usually dominant} if the thermodynamic parameters are modulated although it is zero if the thermodynamic parameters are fixed to zero bias.
\rd{To observe the spin effects, we consider collinear magnetic fields, which relate to spins through the Zeeman effect, 
with different amplitudes applying to the \bl{QDs} and the leads.}
For interacting one level \bl{QD}, we calculate analytically the pumped charge and spin by modulating the magnetic fields and the coupling strengths to the leads 
in the \RD{noninteracting} and strong interacting limi\bl{ts}. 
We show that the difference between these two limits appears through \rd{the \RD{instantaneous} averages of the numbers of the electron with up and down spin in the \bl{QD}}. 
\rd{For the quantum pump by the modulation of the magnetic fields of the \bl{QD} and one lead, the energy dependences of linewidth functions,
 which are usually neglected, are essential.}
\end{abstract}

\maketitle 

\section{Introduction}
In a mesoscopic system, even at zero bias, a charge or spin current is induced by a slow modulation of two or more control parameters\rd{\cite{Thouless83,ex1,ex2,ex3,ex4,ex5,ex6,ex7,ex8,ex9,ex10}}.
This phenomenon, \rd{called} quantum adiabatic pump, is theoretically interesting because its origins are quantum effects and non-equilibrium effects.
The quantum adiabatic pump is also expected to be applied to the single electron transfer devices and the current standard\RD{\cite{CS1,CS2}}.

The \RD{adiabatically} pumped \BL{quantity} is described by a geometric expression in the control parameter space, \RD{although the pumped quantity coming from second or more higher order of the pumping frequency is not geometric}. 
In noninteracting systems, the quantum adiabatic pump had extensively been studied by the Brouwer formula\cite{Brouwer98,BrouwerGeo,BrouwerFCS1,BrouwerFCS2,Brouwer02,Buiittiker02,Shutenko,Wei,Wohlman1},
which describes the pumped charge by the scattering matrix. On the other han\bl{d, i}t is difficult to calculate the scattering matrix in the interacting systems.
In the interacting \bl{system}, the Brouwer formula had only been applied in mean field treatments \cite{aono03,aono04} or in the Toulouse limit \cite{Schiller}.

Recently, the quantum pump in interacting systems have been actively researched. 
There are \rd{three} theoretical approaches. \rd{The first is the Green's function approach to pumping\cite{G1,G2,G3}}.
\rd{The second} is the real-time diagrammatic approach 
\rd{\cite{Splettstoesser06,Konig09-,a0,Splettstoesser10-1,Splettstoesser10-2,Splettstoesser12,Splettstoesser13,Konig13-2} (\rd{RT approach})} 
which uses \rd{the generalized master equation (GME) that is equivalent\cite{a9,a10} to the quantum master equation (QME) derived using the Nakajima-Zwanzig projection operator technique
\cite{open}}. 
Particularly, \Ref{Splettstoesser12} \rd{derived} a geometric expression similar to the Brouwer formula and the \rd{Berry-Sinitsyn-Nemenman (BSN) vector explained later}.
The \rd{third} is the \rd{full counting statistics\cite{FCS-QME,FCS,Utumi} (FCS)  with quantum master equation  (FCS-QME, which is also called generalized quantum master equation\cite{FCS-QME}}) approach 
proposed in \Ref{Main}.

\rd{The adiabatic modulation of the control parameters induces a Berry-phase-like\cite{Berry} quantity called BSN phase in the FCS-QME with the Markov approximation.} 
Sinitsyn and Nemenman\cite{Sinitsyn} \rd{studied} the adiabatically pumped charge using the FCS and had shown that it is 
characterized by the \rd{BSN} vector, \bl{which results from the BSN phase}. 
The BSN vector was applied to the spin boson system \cite{Spin-boson} and a \rd{connection was made} to the excess entropy production\cite{Sagawa,yuge1}.
\rd{The \rd{FCS-QME} approach can treat the Coulomb interaction, which can not be treated in the Brouwer formula}. 
The derived formula of \bl{the BSN vector} depends on the approximation\bl{s} used for the QME. 
The Born-Markov approximation with or without the rotating wave approximation \cite{open}(RWA)
is frequently used. 
The QME in the Born-Markov approximation without RWA sometimes violates the non-negativity of the system reduced density operator \cite{Mar}.
The  QME of the RWA or the coarse-graining approximation\cite{CG,CG13}(CGA) is the Lindblad type which guarantees the non-negativity \cite{open}.

\rd{Some recent papers\cite{Splettstoesser10-2,Splettstoesser12,Main} showed that the Coulomb interaction induces the quantum pump. 
In Refs. [\onlinecite{Splettstoesser10-2,Splettstoesser12}], it was shown that in a one level interacting quantum dot (QD) weakly coupled to two leads, 
the pumped charge (also spin in \Ref{Splettstoesser12}) induced by an adiabatic modulation of the energy level of the QD and the bias between the two leads 
vanishes in the noninteracting limit. 
In particular,} Yuge {\it et al.}\cite{Main} studied the pumped charge coming from the BSN curvatures by adiabatic modulation of the thermodynamic parameters 
(the chemical potentials and the temperatures) in spinless \bl{QDs} weakly coupled to two spinless leads and showed that the BSN curvatures 
are zero in noninteracting \bl{QDs} although they are nonzero for finite interaction.

In this paper, we first generalize the \rd{FCS-QME} approach to multicounting \BL{field} to calculate spin current (\res{FCS-QMEd}). 
We then study the nonadiabatic effects in \rd{general Markov systems} and \bl{clarified the}
 relations between the \rd{FCS-QME} approach and the \rd{RT approach} \cite{Splettstoesser12} in \res{Nona}. 
Additionally we show that the origin of the BSN phase is \rd{a nonadiabatic effect}. 
Next we explain the model to be considered (\res{model}). We consider \bl{QD}s weakly coupled to two leads ($L$ and $R$). 
\rd{To observe the spin effects, we consider collinear magnetic fields, which relate to spins through the Zeeman effect, 
with different amplitudes applying to the \bl{QD}s ($B_S$) and the leads ($B_L$ and $B_R$)}. 
The dynamic parameters (\rd{$B_S$, $B_{L/R}$}, and the coupling strengths to the leads) are control parameters.
We use the RWA defined as a long coarse-graining time limit of the CGA to the \rd{FCS-QME}. 
In \res{Non-interacting} and \res{Interacting}, we consider noninteracting and interacting \bl{QDs} respectively. 
First, we show (in \res{SC} and \res{current,inf}) the time integral of an \rd{instantaneous} steady current is \rd{usually dominant} if the thermodynamic parameters 
(the chemical potentials and the temperatures of leads) are modulated 
(as considered in Refs. [\onlinecite{Main,Splettstoesser10-2,Splettstoesser12,Utiyama, Watanabe}]).
\BL{
Next in} a one level \bl{QD} with the Coulomb interaction $U$, we analytically calculate the BSN curvatures of spin and charge induced by the dynamic parameters 
in the \RD{noninteracting} (\res{pump,U=0}) and strong (\res{pump}) interacting limit ($U\to 0,\infty$).
The difference between the results for $U=0$ and $U=\infty$ appears through the \rd{\RD{instantaneous} averages of the numbers of the electron with up and down spin in the \bl{QD}}. 
\rd{For the quantum pump by the adiabatic modulation of $(B_L, B_S)$, the energy dependences of linewidth functions, which are usually neglected, are essential.}
In \res{pump}, we show and discuss the contour plots of BSN curvatures evaluated numerically.  
\rd{Finally}, we summarize this paper with discussions (\res{Con}). 
In Appendix \ref{Liouville space}, the Liouville space\cite{Fano,FCS-QME} and the matrix representation of the Liouvillian are explained. 
In Appendix \ref{CGA}, we derive the \rd{FCS-QME} of the CGA and discuss the difference between the CGA and the RWA. 
\rd{In Appendix \ref{co}, we derive \re{exp}.}
\rd{In Appendix \ref{val}, we discuss the validity of the adiabatic expansion in \res{Nona}.}
\bl{In Appendix \ref{proof}, we discuss the derivation of \re{W_R}.} 

\section{FCS-QME}

In this section, we consider \rd{general Markov systems} weakly coupled to noninteracting (fermionic or bosonic) baths. 
The model we use to \rd{do} a concrete calculation is explained at \res{model}.

In \res{FCS-QMEd}, we explain the \rd{FCS-QME} method using the Liouville space\cite{Fano,FCS-QME}(Appendix \ref{Liouville space}). 
This method is a generalization to the multicounting \BL{field} of \Ref{Main}. 
In \res{Nona}, we study nonadiabatic effect, and show the equivalence to the method of \Ref{Splettstoesser12}.

\subsection{Derivation of \rd{FCS-QME}} \la{FCS-QMEd}

Consider a cyclic modulation of the control parameters with a period $\tau$. 
At $t=0$ and $t=\tau$, we perform projection measurements of $\mu$th time-independent observables $\{O_\mu \}$ \rd{indexed by $\mu$} of baths
which commute with each other. 
$\Dl o_\mu= o_\mu^{(\tau)}- o_\mu^{(0)} $ denotes the difference of the outcomes $\{ o_\mu^{(\tau)} \}$ at $t=\tau$ and the outcomes $\{ o_\mu^{(0)} \}$ at $t=0$. 
The Fourier transform of 
the joint probability density distribution $P_\tau(\{\Dl o_\mu \})$, $Z_\tau(\{ \chi_\mu \})=\int \prod_\mu d\Dl o_\mu \ P_\tau(\{\Dl o_\mu \})e^{i\sum_\mu \chi_\mu \Dl o_\mu}$, 
is the generating function. 
Here, $\chi_\mu$ are counting fields for $O_\mu$. 
$Z_\tau(\{ \chi_\mu \})$ is given by $Z_\tau(\{ \chi_\mu \})=\tr_\tot[\rho_\tot^\chi(t=\tau)]$ using an operator of \bl{the total system} $\rho_\tot^\chi(t)$\cite{FCS-QME}. 
Here, $\chi$ denotes the set of the counting fields $\{ \chi_\mu \}$. 
We defined $\rho^\chi(t)\defe \tr_B[\rho_\tot^\chi(t)]$ where $\tr_B$ denotes a trace over \bl{baths' degrees} of freedom. 
$\rho^\chi(t)$ provides the generating function $Z_\tau(\{ \chi_\mu \})=\tr_S[\rho^\chi(t=\tau)]$. 
In Appendix \ref{CGA}, we derive the \rd{full counting statistics with quantum master equation (FCS-QME}) 
[\rd{i.e., the equation of motion of $\rho^\chi(t)$}] from the equation of motion of $\rho_\tot^\chi(t)$. 
\rd{In this paper, we set $\hbar=1$.  
We suppose $\rho_\tot(0)=\rho(0) \otimes \rho_B(\al_0)$ and $\rho_\tot(t)\approx \rho(t) \otimes \rho_B(\al_t)$ $(0<t\le \tau)$ where $\rho_\tot(t)$ is the total system state, 
$\rho(t)=\tr_B[\rho_\tot(t)]$ is the system reduced density operator, $\rho_B(\al_t)$ is a tensor product of the grand canonical or canonical distributions of the baths, 
and $\al_t$ is the value of the set of the control parameters at time $t$.  
If the baths are electric leads, $\rho_B(\al_t)$ is given by \re{canon}.}
The \rd{FCS-QME\cite{FCS-QME,Main}} is 
\bea
\f{d \rho^\chi(t)}{dt}=\Hat{K}^\chi(\al_t)\rho^\chi(t), \la{FCS-QME}
\eea
\rd{and the initial condition is $\rho^\chi(0)=\rho(0)$. 
Here $\Hat{K}^\chi(\al_t)$ is the Liouvillian modified by $\chi$. 
The Liouvillian depends on used approximations, for instance, the Born-Markov approximation without or within RWA\cite{open} and the CGA\cite{CG,CG13}. 
After \res{model} we choose the Born-Markov approximation within RWA; however, in this section we assume only Markov property (i.e., $\Hat{K}^\chi$ just depends on $\al_t$).} 
At $\chi=0$, the \rd{FCS-QME} becomes the quantum master equation (QME)
\bea
 \f{d\rho(t)}{dt} = \Hat{K}(\al_t)\rho(t). \la{QME}
\eea
$\Hat{K}(\al_t)$ equals $\Hat{K}^\chi(\al_t)$ at $\chi=0$. In the following, a symbol $X$ without $\chi$ denotes $X^\chi \vert_{\chi=0}$.

In the Liouville space (Appendix \ref{Liouville space}), the left and right eigenvalue equations of the Liouvillian are
\bea
\Hat{K}^\chi(\al)\dke{\rho_n^\chi(\al)}\aeq \lm_n^\chi(\al)\dke{\rho_n^\chi(\al)} \la{rig} ,\\
\dbr{l_n^\chi(\al)}\Hat{K}^\chi(\al)\aeq \lm_n^\chi(\al)\dbr{l_n^\chi(\al)} \la{left}.
\eea
The left eigenvectors $l_n^\chi(\al)$ and the right eigenvectors $\rho_m^\chi(\al)$ 
(\rd{operators considered as elements of a vector space}) satisfy $\dbra l_n^\chi(\al) \dke{\rho_m^\chi(\al)}=\dl_{nm}$. 
Here, $\al$ denotes arbitrary values of the set of the control parameters. 
\bl{The mode which has the eigenvalue with the maximum real part is assigned by the label $n=0$ and is called the slowest mode.}  
\rd{In the limit} $\chi \to 0$, $\lm_0^\chi(\al)$ becomes $0$ and $\dbr{l_0^\chi(\al)}$ becomes $\dbr{1}$, i.e., $l_0(\al)=1$. 
The conservation of the probability $\f{d}{dt}\dbra 1 \dke{\rho(t)}=\dbr{1}\Hat{K}(\al_t)\dke{\rho(t)}=0$ leads
$\dbr{1}\Hat{K}(\al)=0$. 
In addition, $n=0$ mode \rd{right eigenvector, $\dke{\rho_0(\al)}$,} determined by $\Hat{K}(\al)\dke{\rho_0(\al)}=0$ represents the \rd{instantaneous} steady state;  
if the control parameters are fixed to $\al$, the state $\rho(t)$ converges to $\rho_0(\al)$ at $t \to \infty$.
In general, the solution of the \rd{FCS-QME} \re{FCS-QME} is expanded as
\bea
\dke{\rho^\chi(t)} \aeq \sum_n c_n^\chi(t)e^{\Lm_n^\chi(t)}\dke{\rho_n^\chi(\al_t)} \la{exp},
\eea
where $\Lm_n^\chi (t)=\int_0^t ds \ \lm_n^\chi(\al_s)$. The coefficients $c_n^\chi(t)$ obey \rd{\re{comment2}}. 

\RD{The coefficients $c_n^\chi(t)$ are given by solving \re{comment2}. 
The condition which makes $\{c_n^\chi(t) \}_{n \ne0}$ negligible is discussed in Appendix \ref{co} and was studied carefully in \Ref{Lidar05} for $\chi=0$}. 
In this section, we consider sufficiently slow modulation of the control parameters. 
The effects of fast modulation \bl{are} considered in the next section. 
For \bl{the slowest mode}, the second term of the right side of \re{comment2} exponentially damps as a function of time. The relaxation time of the system ($\tau_S$) is the order of $\Ga^{-1}$ where $\Ga$ is \bl{the typical 
value} of the linewidth functio\bl{ns} [defined as $\Ga_i$ in \re{I_Ls}]. 
Assuming the cycle time $\tau$ is much \bl{longer} than $\tau_S$, we obtain 
\bea
c_0^\chi(\tau) \aeq c_0^\chi(0)\exp \Big[-\int_0^\tau dt \  \dbr{l_0^\chi(\al_t)} \f{d }{dt}\dke{\rho_0^\chi(\al_t)} \Big] \la{BSN_p},
\eea
and 
\rd{\bea
c_n^\chi(\tau)e^{\Lm_n^\chi(\tau)} \approx 0 \ (n \ne 0) .\la{BSN_p2} 
\eea
In fact, $c_n^\chi(t)e^{\Lm_n^\chi(t)}=\mO(\f{\om}{\Ga})$ with $\om=2\pi/\tau$ as explained in Appendix \ref{co}.
We denote the condition which makes the above approximation appropriate as {\it $\chi$-adiabatic condition. }}
 Using the initial condition $\rho^\chi(0)=\rho(0)$, we obtain $c_0^\chi(0)=\dbra l_0^\chi(\al_0)\dke{\rho(0)}$. 
Substituting these equations into \re{exp}, we obtain \cite{Main} 
\bea
\dke{\rho^\chi(\tau)} 
\aeqap \dbra l_0^\chi(\al_0) \dke{\rho(0)}  e^{ -\int_0^\tau dt \ \dbr{l_0^\chi(\al_t)}\f{d}{dt} \dke{\rho_0^\chi(\al_t)}}\no\\
&&\times e^{\int_0^\tau dt \ \lm_0^\chi(\al_t) }\dke{\rho_0^\chi(\al_\tau)} , \la{sol_ap}
\eea
and the cumulant generating function $S_\tau(\chi)=\ln Z_\tau(\chi)=\ln \dbra 1 \dke{\rho^\chi(\tau)}$ :
\bea
S_\tau(\chi) \aeq \int_0^\tau dt \ \lm_0^\chi(\al_t)-\oint_C d\al^n \  \dbr{l_0^\chi(\al)} \f{\partial \dke{\rho_0^\chi(\al)}}{\partial \al^n} \no\\
&&+\ln \dbra l_0^\chi(\al_0)\dke{\rho(0)} +\ln \dbra 1 \dke{\rho_0^\chi(\al_\tau)}. \la{S_tau}
\eea
Here, we used $\int_0^\tau dt \  \dbr{l_0^\chi(\al_t)} \f{d }{dt}\dke{\rho_0^\chi(\al_t)}=\oint_C d\al^n \  \dbr{l_0^\chi(\al)} \f{\partial \dke{\rho_0^\chi(\al)}}{\partial \al^n}$, 
where $C$ is the trajectory from $\al_0$ to $\al_\tau$, $\al^n$ are the $n$th component of the control parameters and the summation symbol $\sum_{n}$ is omitted. 
\RD{Equation (\ref{S_tau}) is the same with Yuge {\it et al.}\cite{Main} except for that $\chi$ denotes a multicounting field. }
The averages $\bra \Dl o_\mu \ket_\tau= \f{\partial S_\tau(\chi) }{\partial (i\chi_\mu)} \big \vert_{\chi=0}$ are 
\bea
\bra \Dl o_\mu \ket_\tau 
\aeq \int_0^\tau dt \ \lm_0^\mu(\al_t)-\oint_C d\al^n \  \dbr{l_0^\mu (\al)} \f{\partial \dke{\rho_0(\al)}}{\partial \al^n} \no\\
&&+\dbra l_0^\mu(\al_0)\dke{\rho(0)}+\dbra 1 \dke{\rho_0^\mu(\al_0)} ,\la{ave}
\eea
\rd{where $X^\mu(\al)\defe \f{\partial X^\chi(\al) }{\partial (i\chi_\mu)} \big \vert_{\chi=0}$. 
Here, we used $-\oint_C d\al^n \ \dbr{l_0(\al)} \f{\partial \dke{\rho_0^\mu(\al)}}{\partial \al^n}=-\dbra 1 \dke{\rho_0^\mu(\al_\tau)}+\dbra 1 \dke{\rho_0^\mu(\al_0)} $ 
because $\dbr{l_0(\al)} \f{\partial \dke{\rho_0^\mu(\al)}}{\partial \al^n}= \f{\partial }{\partial \al^n}\dbra 1 \dke{\rho_0^\mu(\al)}$. 
The integrand of the first time integral, $\lm_0^\mu(\al_t)$,}  are the \rd{instantaneous} steady currents of $O_\mu$ at time $t$;
if the control parameters are fixed to $\al$ and the state is $\rho_0(\al)$, the current of $O_\mu$ is $\lm_0^\mu(\al)$.
The integrand of the second term of the right side of \re{ave} is the Berry-Sinitsyn-Nemenman (BSN) vector \cite{Sinitsyn} 
\bea
A_n^\mu(\al)=\dbr{l_0^\mu (\al)} \f{\partial \dke{\rho_0(\al)}}{\partial \al^n} \la{BSNv}.
\eea
The third and fourth terms of the right side of \re{ave} cancel if the initial condition is the \rd{instantaneous} steady state $\rho_0(\al_0)$.
Because of $\al_\tau=\al_0$, the second term of the right side of \re{ave} can be described as a surface integral over the surface $S$ enclosed by $C$ using the Stokes theorem :
\bea
\bra \Dl o_\mu \ket_\tau \aeq \bra \Dl o_\mu \ket^{\rm{Steady}}_\tau+\bra \Dl o_\mu \ket^{\rm{Berry}}_S ,\la{yuge} \\
\bra \Dl o_\mu \ket^{\rm{Steady}}_\tau \aeq \int_0^\tau dt \ \lm_0^\mu(\al_t) \la{yuge_S},\\
\bra \Dl o_\mu \ket^{\rm{Berry}}_S \aeq -\int_S d\al^m \wedge d\al^n \ \half F^\mu_{mn}(\al) \la{yuge_B}  .
\eea
Here, $\wedge$ is the wedge product and the summation symbol $\sum_{n,m}$ is omitted. BSN curvature $F^\mu_{mn}(\al)$ is given by 
\rd{\bea
F^\mu_{mn}(\al)\aeq \f{\partial A_n^\mu(\al)}{\partial \al^m}-\f{\partial A_m^\mu(\al)}{\partial \al^n}. 
\eea
Yuge {\it et al.}\cite{Main} focus on only the second term of \re{yuge} subtracting the first term, and they did not evaluate} $\bra \Dl o_\mu \ket^{\rm{Steady}}_\tau$.
In \res{SC}, we show that this contribution is \rd{usually dominant} if the thermodynamic parameters are modulated 
although the steady currents $\lm_0^\mu(\al_t)$ are zero if the thermodynamic parameters are fixed to zero bias.

\rd{From the cumulant generating function \re{S_tau}, we can calculate the second order cumulants 
${}^c\bra o_\mu o_\nu \ket_\tau= \f{\partial^2 S_\tau(\chi) }{\partial (i\chi_\mu)\partial (i\chi_\nu)} \big \vert_{\chi=0}$ and higher cumulants.
However, we focus only on the first order cumulants (averages) in this paper. Up to the  first order cumulants, we do not need the multicounting \BL{field}. 
In fact, the multicounting \BL{field} is helpful to understand the physical origin of each term of the (FCS-)QME.} 
\RD{The counting fields appear only in $A \bu B$ ($A,B \ne 1$) type terms in the dissipator term $\Hat{\Pi}^\chi\bu $ [see \re{K}; $\bu$ is an arbitrary operator] as \re{CG_go}. 
For instance, the factor $e^{i\chi_{b\sig}}$ with the counting field for total number of electron with spin $\sig$ of bath $b$, $\chi_{b\sig}$, means tunneling process from the system to bath $b$ with spin $\sig$. 
The factor $e^{-i\chi_{b\sig}}$ means tunneling process from bath $b$ with spin $\sig$ to the system.}

\subsection{Non-adiabatic effect and BSN vector} \la{Nona}

In this section, we consider nonadiabatic effect \rd{[that comes from $\big(\f{\om}{\Ga}\big)^{n}\ (n=1,2,\cdots)$ with $\om=2\pi/\tau$]}, 
which had been researched recently\rd{\cite{a0,Splettstoesser13, Utiyama, Watanabe}}. 
If the modulation of the control parameters \bl{is} not adiabatic, the difference between the state and the \rd{instantaneous} steady state, 
$\rho^a(t) \defe \rho(t)-\rho_0(\al_t) =\sum_{n\ne 0}c_n(t)e^{\Lm_n(t)}\rho_n(\al_t)$, is important. 
\RD{In contrast to Appendix \ref{co} and \Ref{Lidar05}}, we do not treat $\{c_n(t) \}_{n \ne0}$ explicitly; instead, we use the pseudoinverse of the Liouvillian.
The formal solution of the \rd{FCS-QME} \re{FCS-QME} is
\bea
\dke{\rho^\chi(t)} \aeq {\rm{T}} \exp \Big[\int_0^t ds \ \Hat{K}^\chi(\al_s) \Big]\dke{\rho^\chi(0)} ,
\eea
where $ {\rm{T}}$ denotes the time-ordering operation. Using this, we obtain the averages \cite{Utiyama}
\bea
\bra \Dl o_\mu \ket_t \aeq \f{\partial }{\partial(i\chi_\mu) } \dbra 1 \dke{\rho^\chi(t)}   \Big \vert _{\chi=0} \no\\
\aeq \int_0^t du \ \dbr{1}  \Hat{K}^\mu(\al_u)\dke{\rho(u)}+ \dbra 1\dke{\rho^\mu(0)} \no\\
\aeq  \int_0^t du \ \dbr{1}\Hat{K}^\mu(\al_u)\dke{\rho(u)} \e  \int_0^t du \  I_\mu (u). \la{wat}
\eea
Here, we use $\dbr{1}\Hat{K}(\al)=0$ in the second line and $\dke{\rho^\mu(0)}=0 $ (derived from $\dke{\rho^\chi(0)}=\dke{\rho(0)}$) in the third line.
Moreover, using $\dbr{l_0(\al)}=\dbr{1}$, $\lm_0(\al)=0$ and \re{left}, we obtain\cite{Watanabe}
\bea
\dbr{1}\Hat{K}^\mu(\al)
\aeq \lm_0^\mu (\al) \dbr{1} - \dbr{l_0^\mu (\al)}\Hat{K}(\al). \la{wata}
\eea
\bl{Hence} the currents $I_\mu (t)$ are given by 
\bea
 I_\mu (t) \aeq \dbr{1}\Hat{K}^\mu(\al_t)\dke{\rho(t)} \no\\
\aeq \lm_0^\mu (\al_t) - \dbr{l_0^\mu (\al_t)}\Hat{K}(\al_t) \dke{\rho(t)}  \no\\
\aeq \lm_0^\mu (\al_t)- \dbr{l_0^\mu (\al_t)}\f{d}{dt}\dke{\rho(t)}. \la{watan}
\eea
\rd{The used approximation is only the Markov property of the Liouvillian, \re{FCS-QME}. 
In Appendix \ref{val}, we discuss the reasonable range of the power of $\f{\om}{\Ga}$ (nonadiabaticity).}
Substituting \rd{\re{watan}} with $\rho(t) \approx \rho_0(\al_t)$ into \re{wat}, we obtain \re{ave} without the third and fourth terms. 
If $\rho(0)\ne \rho_0(\al_0)$, the state \rd{relaxes} to the \rd{instantaneous} steady state in the relaxation time $\tau_S$. 
The third and fourth terms of \re{ave}\rd{, $\bra \Dl o_\mu \ket_\tau^{3+4}=\dbra l_0^\mu(\al_0)\dke{\rho(0)}+\dbra 1 \dke{\rho_0^\mu(\al_0)}$, 
result from this relaxation.  
The contribution of $\bra \Dl o_\mu \ket_\tau$ from $\dl \rho(0)=\rho^a(0)=\rho(0)-\rho_0(\al_0)$ is 
\bea
\bra \Dl o_\mu \ket_\tau^{\rm{ini}} \aeqd -\int_0^\tau dt \ \dbr{l_0^\mu (\al_t)}\f{d}{dt}\dke{\dl \rho(t)} \no\\
\aeq \dbra l_0^\mu (\al_0)\dke{\dl \rho(0)}-\dbra l_0^\mu (\al_\tau)\dke{\dl \rho(\tau)}\no\\
&&+\int_0^\tau dt \ \f{d\dbr{l_0^\mu (\al_t)}}{dt}\dke{\dl \rho(t)}, \la{con_ini}
\eea
with $\dke{\dl \rho(t)}\defe{\rm{T}} \exp \Big[\int_0^t ds \ \Hat{K}(\al_s) \Big]\dke{\dl \rho(0)}$. 
The first term of the right side of \re{con_ini} is $\bra \Dl o_\mu \ket_\tau^{3+4}$. 
Because we can obtain $\dbr{l_0^\mu(\al)} \rho_0(\al) \dket+\dbr{1} \rho_0^\mu(\al) \dket=0$ from the normalization $\dbra l_0^\chi(\al) \dke{\rho_0^\chi(\al)}=1$, 
$\bra \Dl o_\mu \ket_\tau^{3+4}$ is given by $\dbra l_0^\mu (\al_0)\big[\dke{\rho(0)}-\dke{\rho(\al_0)} \big]=\dbra l_0^\mu (\al_0)\dke{\dl \rho(0)}$. 
The second term of the right side of \re{con_ini} is exponentially small since $\dl \rho(\tau) \sim e^{-\Ga \tau}$. 
The order of the third term is $\mO(\f{\om}{\Ga})$ with $\om=2\pi/\tau$ because $\f{d\dbr{l_0^\mu (\al_t)}}{dt}=\mO(\om)$ and the integral range is restricted up to $1/\Ga$ 
since $\dl \rho(t) \sim e^{-\Ga t}$. Hence $\bra \Dl o_\mu \ket_\tau^{\rm{ini}}=\bra \Dl o_\mu \ket_\tau^{3+4}+\mO(\f{\om}{\Ga})$. }

\rd{The currents can also be written as}
\bea
 I_\mu(t) 
\aeq \dbr{1} W_\mu(\al_t) \dke{\rho(t)} , \la{I_mu}
\eea
where $W_\mu(\al)$ are the current operators defined by \rd{
\bea
\dbr{1}W_\mu(\al) = \dbr{1}\Hat{K}^\mu(\al) \la{defW},
\eea
i.e., $\tr_S[W_\mu(\al) \bu] =\tr_S[\Hat{K}^\mu(\al)\bu]$ for any operator $\bu$ [see, \re{W_AB}].}
Using \re{wata}, the \rd{instantaneous} steady currents are given by 
\bea
\dbr{1}W_\mu (\al) \dke{\rho_0(\al)} \aeq \lm_0^\mu(\al)=I_\mu^{\rm{Steady}}(\al) . \la{touka_sc}
\eea
In the \bl{QDs} weakly coupled to two leads, the electric current operator [i.e., $W_\mu(\al)$ corresponding to the electric current] coincides with \Ref{Yan05} 
in the Born-Markov approximati\bl{on}
 without or within RWA. 

\bl{Applying the pseudoinverse $\mR(\al)$ defined by 
\bea
\mR(\al)\Hat{K}(\al)=1-\dke{\rho_0(\al)}\dbr{1}, \la{defR}
\eea
to the QME \re{QME}, we obtain}
\bea
\dke{\rho^a(t)} \aeq  \mR(\al_t) \f{d}{dt}\dke{\rho_0(\al_t)}+\mR(\al_t)\f{d}{dt}\dke{\rho^a(t)} \no\\
\aeq \sum_{n=1}^\infty \Big[ \mR(\al_t) \f{d}{dt} \Big]^n \dke{\rho_0(\al_t)} \e \sum_{n=1}^\infty \dke{\rho^{a(n)}(t)}. \la{key6}
\eea
\rd{Substituting} \re{key6} to \re{I_mu}, we finally reach
\bea
I_\mu (t) =I_\mu^{\rm{Steady}}(\al_t) + \sum_{n=1}^\infty I_\mu^{a(n)}(t) \la{key7},
\eea
\bl{with $I_\mu^{a(n)}(t) \defe\dbr{1} W_\mu(\al_t)\dke{\rho^{a(n)}(t)} $}. 
\rd{Since $\f{d}{dt}\al_t=\mO(\om)$ and $\mR(\al_t)=\mO(\f{1}{\Ga})$,
\bea
\rho^{a(n)}(t)=\mO \big(\f{\om}{\Ga} \big)^n. \la{order}
\eea
In Appendix \ref{val}, we discuss the reasonable range of $n$ of $\rho^{a(n)}(t)$ and show that with the larger nonadiabaticity ($\f{\om}{\Ga}$), the reasonable range becomes wider.}

Let's consider the relation between \re{watan} and \re{key7}. In \res{FCS-QMEd}, we used \rd{$\chi$-adiabatic approximation} \re{sol_ap}, which becomes $\dke{\rho(t)} \approx  \dke{\rho_0(\al_t)}$ at $\chi=0$. 
\rd{Substituting} it to \re{I_mu}, we obtain 
$ I_\mu (t) \approx  I_\mu^{\rm{Steady}}(t) $. So, we cannot obtain nonadiabatic currents $ \sum_{n=1}^\infty I_\mu^{a(n)}(t)$. 
However, from the $\chi_\mu$ derivative of \re{sol_ap}, we obtain
\bea
 I_\mu (t) \aeqap \lm_0^\mu(\al_t)-\dbr{l_0^\mu(\al_t)}\f{d}{dt} \dke{\rho_0(\al_t)} \la{key8}.
\eea
This is equivalent to \re{ave} for $\rho(0)=\rho_0(\al_0)$. 
Equation (\ref{key8}) \bl{suggests}
\bea
I_\mu^{a(1)}(t)= -\dbr{l_0^\mu(\al_t)}\f{d}{dt} \dke{\rho_0(\al_t)}  \la{I^a} .
\eea
\bl{In fact, t}his is equivalent to \bl{$\RD{I_\mu^{a(1)}(t)}=\dbr{1} W_\mu(\al_t)\dke{\rho^{a(1)}(t)}$, namely} 
\bea
I_\mu^{a(1)}(t)=\dbr{1} W_\mu(\al_t)\mR(\al_t)\f{d}{dt}\dke{\rho_0(\al_t)}  \la{key9},
\eea
\bl{because of} 
\bea
\dbr{1} W_\mu(\al) \mR (\al)=-\dbr{l_0^\mu(\al)}+c_\mu(\al) \dbr{1}, \la{W_R}
\eea
\bl{which was shown by Sagawa, {\it et al.} \cite{Sagawa} for a single counting field. Here, $c_\mu(\al)$ are constants shown in \re{Y2}. 
We prove \re {W_R} at Appendix \ref{proof}. }
\bl{Equation}(\ref{watan}) and \re{key7} are identical because of \re{touka_sc} and \re{W_R}. 
\bl{In other words, in the expansion of} $I_\mu(t)$ obtained from a substitution of $\rho(t)=\rho_0(\al_t)+\sum_{n=1}^\infty \rho^{a(n)}(t)\e \sum_{n=0}^\infty \rho^{a(n)}(t) $ 
into \re{watan}, the 
$n$th $(n=0,1,\cdots)$ order nonadiabatic solution, $\rho^{a(n)}(t)$,  gives $(n+1)$th order  
 nonadiabatic currents $I_\mu^{a(n+1)}(t)$ because of \re{W_R}. 
\bl{Hence} the \rd{FCS-QME} approach picks out one higher order nonadiabatic piece of information from the solution of the QME. 
 
 \rd{\bl{Moreover,} although the BSN phase [i.e., the argument of the exponential function of \re{BSN_p}] is derived under 
the $\chi$-adiabatic condition which makes Eqs. (\ref{BSN_p}) and (\ref{BSN_p2}) appropriate, its origin is probably a 
nonadiabatic effect that \RD{comes from $\f{\om}{\Ga}$},} 
because \re{I^a} shows that the BSN phase has the information of \bl{the} nonadiabatic part of the QME $[\rho^a(t)=\rho(t)-\rho_0(\al_t)]$.

It is important to \bl{recognize the} relations between the \rd{FCS-QME} approach and the \rd{RT approach} 
\cite{Konig09-,a0,Splettstoesser10-1,Splettstoesser10-2,Splettstoesser12,Splettstoesser13,Konig13-2}.
In the \rd{RT approach, $p_\ka(t)=\br{\ka}\rho(t)\ke{\ka}$} are governed by the generalized master equation (GME)
\rd{\bea
\f{d}{dt} p_\ka(t) \aeq \sum_\eta \int_{-\infty}^t dt^\pr \ W_{\ka \eta}(t,t^\pr)p_\eta(t^\pr), \la{GME}
\eea
where $\ke{\ka}$} are \RD{the energy eigenstates of the system Hamiltonian}. The kernel $W_{\ka \eta}(t,t^\pr)$ can include the higher order contribution of the tunneling interaction between baths 
and the system. 
In the GME, $p_\eta(t^\pr)$ is given by \rd{$p_\eta(t)+\sum_{k=1}^\infty\f{(t^\pr-t)^k}{k!}\f{d^kp_\eta(t)}{dt^k}$ \cite{Konig09-,a0}}. 
Moreover, $W_{\ka \eta}(t,t^\pr)$ and $p_\eta(t)$ are expanded as $W_{\ka \eta}(t,t^\pr)=\sum_{n=0}^\infty \BL{\sum_{j=1}^\infty W_{\ka \eta(j)}^{(n)}(t;t-t^\pr)}$ 
and $p_\eta(t)=\sum_{n=0}^\infty \BL{\sum_{j=-n}^\infty p_{\eta(j)}^{(n)}(t)}$, where 
$W_{\ka \eta\BL{(j)}}^{(n)}(t;t-t^\pr)$ and $p_{\eta\BL{(j)}}^{(n)}\BL{(t)}$ are of the order of $\om^n\BL{\Ga^j}$. 
In particular, $W_{\ka \eta\BL{(j)}}^{(0)}(t;t-t^\pr)=W_{\ka \eta\BL{(j)}}^{(0)}(\al_t;t-t^\pr)$ is the kernel 
where the control parameters are fixed to $\al_t$.
Up to the second order of the tunneling interaction (in the following we consider this level of approximation), we obtain\cite{a0,Splettstoesser13}
\bea
0 \aeq \sum_\eta K_{\ka \eta}^{(0)}(\al_t) p_{\eta}^{(0)}(\al_t) ,\la{ss_RT}\\
 \f{dp_{\ka\BL{(-n)}}^{(n)}(t)}{dt} \aeq \sum_\eta K_{\ka \eta}^{(0)}(\al_t)p_{\eta\BL{(-n-1)}}^{(n+1)}(t) \la{n_RT},
\eea
for $n=0,1,\cdots$, with
\bea
K_{\ka \eta}^{(0)}(\al_t)=\int_{-\infty}^t dt^\pr \ W_{\ka \eta\BL{(1)}}^{(0)}(\al_t,t-t^\pr),
\eea
which is the \rd{instantaneous} Liouvillian corresponding to our $\Hat{K}(\al_t)$. 
Equation (\ref{ss_RT}) is just the definition of the \rd{instantaneous} steady state 
$p_{\eta}^{(0)}(\al_t)\e p_{\eta\BL{(0)}}^{(0)}(t)$, which satisfies $\sum_\ka  p_{\ka}^{(0)}(\al_t)=1$. 
Additionally, $p_{\ka\BL{(j)}}^{(n)}(t)$ for $n\ge 1$ satisfies $\sum_\ka  p_{\ka\BL{(j)}}^{(n)}(t)=0$.
The conservation of the probability \rd{leads to} $\sum_\ka K_{\ka \eta}^{(0)}(\al_t)=0$, which corresponds to our $\dbr{1}\Hat{K}(\al_t)=0$. 
The charge or spin current $I_\mu(t)$ is given by \cite{Splettstoesser12,Splettstoesser13}
\bea
I_\mu(t)\aeq \sum_{\ka,\eta} w^{[\mu]}_{\ka \eta}(\al_t)p_\eta(t) \la{I_mu_rtd} ,
\eea
\RD{corresponding to our \re{I_mu}. $w^{[\mu]}_{\ka \eta}(\al_t)$ is the instantaneous current matrix of $O_\mu$ in the present approximation, which corresponds to our $W_\mu(\al_t)$ and is linear in $\Ga$.}
Substituting $p_\eta(t)\approx \sum_{n=0}^\infty p_{\eta\BL{(-n)}}^{(n)}(t)$ into \re{I_mu_rtd}, we obtain 
\bea
I_\mu(t)\aeq \sum_{n=0}^\infty I_\mu^{(n)}(t) ,\ I_\mu^{(n)}(t)= \sum_{\ka,\eta} w^{[\mu]}_{\ka \eta}(\al_t)p_{\eta\BL{(-n)}}^{(n)}(t). \la{I_mu^k}
\eea
\rd{Equation (\ref{n_RT}) for $n=0$} leads to \cite{Splettstoesser12}
 \bea
 p_{\eta\BL{(-1)}}^{(1)}(t) \aeq \sum_{\ka} R_{\eta \ka}(\al_t)  \f{dp_{\ka}^{(0)}(\BL{\al_t})}{dt} \la{p^1} .
 \eea
Here, $R_{\eta \ka}(\al_t) $ is the pseudoinverse of $K_{\ka \eta}^{(0)}(\al_t)$ corresponding to our $\mR(\al_t)$ and it is given by \cite{Splettstoesser12}
\bea
R_{\eta \ka}(\al_t)=(\tl K^{-1})_{\eta \ka}, \ \tl K_{\eta \ka}= K_{\eta \ka}^{(0)}- K_{\eta \eta}^{(0)} \la{RR} .
\eea
Substituting \re{p^1} into \re{I_mu^k}, we obtain \cite{Splettstoesser12}
\bea
I_\mu^{(1)}(t)\aeq \sum_{\ka} \varphi_\ka^{[\mu]}(\al_t) \f{dp_{\ka}^{(0)}(\BL{\al_t})}{dt} , \la{I_mu_rtd2} \\
\varphi_\ka^{[\mu]}(\al_t) \aeq \sum_{\zeta, \eta} w^{[\mu]}_{\zeta \eta}(\al_t)R_{\eta \ka}(\al_t) \la{varphi}.
\eea
A similar method has been used in \Ref{Konig09-}. \bl{$\varphi_\ka^{[\mu]}(\al_t)$} and \re{I_mu_rtd2} respectively correspond to our $\dbr{1} W_\mu(\al) \mR (\al)$ and \re{key9}. 
\rd{Moreover, \re{n_RT} for arbitrary $n$ leads to
\bea
 p_{\eta\BL{(-n-1)}}^{(n+1)}(t) \aeq \sum_{\ka} R_{\eta \ka}(\al_t)  \f{dp_{\ka\BL{(-n)}}^{(n)}(t)}{dt} \la{p^n+1} ,
\eea
which corresponds to our \re{key6}.} Because of these relations, the \rd{RT approach} is equivalent to the \rd{FCS-QME} approach in the calculation up to the second order of the tunneling interaction. 
\RD{Additionally, we discuss corrections due to the nonadiabatic effect of the FCS-QME in Appendix \ref{val}. 
The first equation of \re{D1} is consistent with $p_{\eta(0)}^{(1)}(t)=\mO(\om \tau_B)$, which can be derived from \Ref{a0}. Here, $\tau_B$ is the relaxation time of the baths.}

\bl{In this section, we proved the equivalence between \re{watan} and \re{key7} using a key relation \re{W_R} and showed the origin of the BSN phase is a nonadiabatic effect, and
connected} the \rd{FCS-QME} approach and the \rd{RT approach}\cite{Splettstoesser12}. \bl{These are among} the most important results of this paper.

\section{Model} \la{model}

We consider quantum dots \bl{(QDs)} (denoted by \bl{a symbol} $S$) weakly coupled to two leads. 
The total Hamiltonian is \rd{$H_\tot(t)=H_S(t)+\sum_{b=L,R} [H_b(t)+H_{Sb}(t)] $}. 
Here, $H_S(t)$ is the system (QDs) Hamiltonian, $H_b(t)$ is the Hamiltonian of 
lead $b=L,R$, and \rd{$H_{Sb}(t)$} is the tunneling interaction Hamiltonian between $S$ and lead $b$. 
\rd{To observe the spin effects}, we suppose that the leads and the system are applied to collinear magnetic fields with different amplitudes, \rd{which relate to spins through the Zeeman effect}. 
The leads are noninteracting: 
\bea
H_b(t)=\sum_{k,\sig} (\ep_{bk}+\sig g_b B_b(t))c_{bk\sig} \dg c_{bk\sig} . \la{H_b}
\eea
Here, $\sig=\up,\dw=\pm 1$ is spin label, $g_b=\half \mu_{\rm{B}}g_b^\ast $ where $g_b^\ast$ is the $g$-factor of lead $b$, $\mu_{\rm{B}}$ is the Bohr magneton and $B_b(t)$ is the strength of the magnetic field of lead $b$. 
$c_{bk\sig} \dg (c_{bk\sig})$ is the creation (annihilation) operator of an \rd{electron} with spin $\sig$ and momentum $k$ in lead $b$. The system Hamiltonian is
\rd{\bea
H_S(t)=\sum_{n,m,s,s^\pr} \ep_{n s, m s^\pr}(B_S(t)) a_{n s} \dg a_{m s^\pr}+H_{\rm{Coulomb}} , \la{H_S}
\eea
w}here $a_{n s} \dg$ is the creation operator of an \rd{electron} with orbital $n$ and spin $s$. 
$\ep_{n s, m s^\pr}(B_S(t))$ means the energy of the electron for $n=m,s=s^\pr$ and the tunneling amplitu\bl{de} between orbitals for $(n,s)\ne(m,s^\pr)$ 
which depen\bl{ds} on the magnetic field of the system. \rd{$H_{\rm{Coulomb}}$} denotes Coulomb interaction.
The tunneling interaction Hamiltonian is
\rd{\bea
H_{Sb}(t)=\sum_{k,\sig,n,s}\sqrt{\Dl_b(t)}v_{bk\sig,n s}a_{n s}\dg c_{bk\sig}+\hc , \la{H_1}
\eea
where} $\Dl_b(t)$ is a dimensionless parameter, and $v_{bk\sig,n s}$ is the tunneling amplitude.

We assume $B_S$, $B_{L/R}$ and $\Dl_{L/R}$ are control parameters (denoted $\al^\pr=\{B_S,B_{L/R}, \Dl_{L/R}\}$ and are called the dynamic parameters). 
The thermodynamic parameters (the  chemical potentials and inverse temperatures of leads, $\{\mu_b\}$ and $\{\be_b \}$) are also considered as control parameters in \res{SC} and \res{current,inf}. 
We denote $\al^{\pr\pr}=\{\be_b, \mu_b\}_{b=L,R}$ and $\al=\al^\pr+\al^{\pr\pr}$.
Yuge {\it et al.}\cite{Main} chose the set of control parameters as only $\al^{\pr\pr}$. However we are interested in $\al^\pr$ \rd{for the reason} explained in \res{SC}.  

We choose the measured observables $\{ O_\mu \} =\{N_{b\sig} \}_{\sig=\up,\dw}^{b=L,R}$ with $N_{b\sig}=\sum_{k}c_{bk\sig} \dg c_{bk\sig}$. 
The pumped charge (spin) of lead $b$ is given by \rd{$\bra \Dl N_{b\up} \ket\pm\bra \Dl N_{b\dw} \ket$. $\bra \Dl N_{b\sig} \ket$ are calculated by \re{yuge}. 
In fact, what we call the pumped charge, $\bra \Dl N_{b\up} \ket+ \bra \Dl N_{b\dw} \ket$, is the pumped electron number (actual pumped charge is given by $-e[\bra \Dl N_{b\up} \ket+ \bra \Dl N_{b\dw} \ket]$, where $e\ (>0)$ is the elementary charge).}

In
\res{pump,U=0} and \res{Interacting} we consider a one level system
\bea
H_S(t)=\sum_{s=\up,\dw} \om_s(B_S(t)) a_s \dg a_s+Ua_\up \dg a_\up a_\dw \dg a_\dw , \la{H_S,1}
\eea
as a special \bl{model} of \re{H_S}. Here, \bl{$s=\up,\dw=\pm 1$,} $\om_s(B_S)=\om_0+sg_SB_S$ with $\om_0$ the electron energy at $B_S=0$, and $g_S=\half \mu_{\rm{B}}g_S^\ast $ where $g_S^\ast$ is the $g$ factor of the \bl{QD}.

\rd{In the following, we apply the \rd{FCS-QME} with rotating wave approximation (RWA) explained in Appendix \ref{CGA}.}

\section{Non-interacting system} \la{Non-interacting}

In this section, we consider a noninteracting system ($H_{\rm{Coulomb}}=0$). The system Hamiltonian \re{H_S} can be diagonalized
\bea
H_S \aeq \sum_{i=1}^{2N} \tl \om_{i} b_{i} \dg b_{i},
\eea
by a unitary transform $a_{n s} =\sum_{i=1}^{2N} U_{n s,i} b_i $. The tunneling interaction Hamiltonian \re{H_1} is 
\bea
 \RD{H_{Sb}} \aeq \sum_{k ,\sig,i} W_{bk\sig, i} b_{i} \dg c_{bk\sig} +\hc ,
\eea
with $W_{bk\sig,i} =\sum_{n,s} \sqrt{\Dl_b} v_{bk\sig,n s} U_{n s,i} ^\ast $.

In \res{Liouvillian,0}, the Liouvillian and its \rd{instantaneous} steady state are explained. In \res{SC}, we consider the contribution of \re{yuge_S} and show that this cannot be neglected in general if 
the chemical potentials and the temperatures are not fixed.
In \res{pump,U=0}, we calculate the BSN curvatures for two combinations of modulated control parameters $(B_L,B_S)$ and $(\Dl_L,B_S)$.

\subsection{Liouvillian} \la{Liouvillian,0}

The Liouvillian in the RWA is given by
\bea
\Hat{K}^{\chi}(\al ) \aeq \sum_{i=1}^{2N}  \Hat{K}_i^{\chi}(\al) ,\la{L_non} \\
\Hat{K}_i^{\chi}(\al )\bu \aeq -i[\tl \om_i b_i \dg b_i,\bu]+ \Hat{\Pi}_i^{\Phi}(\chi,\al )\bu+ \Hat{\Pi}_i^{\Psi}(\al)\bu ,
\eea
if $\{\tl \om_i \}$ are not degenerated. Here, superoperators $\Hat{\Pi}_i^{\Phi}(\chi,\al )$ and $\Hat{\Pi}_i^{\Psi}(\al)$ operate to an arbitrary operator $\bu$ as   
\bea
&&\hs{-10mm}\Hat{\Pi}_i^{\Phi}(\chi,\al )\bu \no\\
\aeq  \sum_{b}  \Big\{ 
\Phi_{b,i}^{+,\chi} b_i \dg \bu b_i -\half  \Phi_{b,i}^+ \bu b_ib_i \dg 
-\half\Phi_{b,i}^+ b_i   b_i   \dg \bu  \no\\
&&+\Phi_{b,i}^{-,\chi} b_i \bu  b_i \dg 
-\half\Phi_{b,i}^- \bu b_i   \dg b_i -\half \Phi_{b,i}^- b_i \dg b_i \bu   \Big\} ,\\
&&\hs{-10mm}\Hat{\Pi}_i^{\Psi}(\al) \bu = i[ \Om_i(\al)b_i \dg b_i  ,\bu] ,
\eea
with 
\bea
\Phi_{b,i}^{\pm,\chi} \aeq 2\pi \sum_{k,\sig} \abs{W_{bk\sig,i}}^2 f_b^\pm(\tl \om_i) e^{\mp i\chi_{b\sig}} \no\\
&&\hs{10mm}\times \dl(\ep_{bk}+\sig g_bB_b-\tl \om_i), \\
\Psi_{b,i}^{\pm}\aeq 2\sum_{k,\sig} \abs{W_{bk\sig,i}}^2 f_b^\pm(\tl \om_i) {\rm{P}}\f{1}{\ep_{bk}+\sig g_bB_b-\tl \om_i} , 
\eea
and 
$\Om_i(\al)=\half \sum_{b} \Big(\Psi_{b,i}^-+\Psi_{b,i}^+ \Big)$. 
Here, $f_b^+(\om)=[e^{\be_b(\om-\mu_b)}+1]^{-1}$ is the Fermi distribution function, $f_b^-(\om)=1-f_b^+(\om)$, $\chi_{b\sig}$
 is the counting field for $N_{b\sig}$ and $\rm{P}$ denotes the Cauchy principal value.
The matrix representation of $\Hat{K}_i^\chi(\al)$ (see Appendix \ref{Liouville space}) by the number states of $b_i \dg b_i$ ($\ke{0}_i$ and $\ke{1}_i$) is a $4\times 4$ matrix which is block diagonalized to $\{ \ke{0}_i {}_i\br{0}, \ke{1}_i {}_i\br{1} \}$ space and 
$\{ \ke{0}_i {}_i\br{1}, \ke{1}_i {}_i\br{0} \}$ space. The $\{ \ke{0}_i {}_i\br{0}, \ke{1}_i {}_i\br{1} \}$ part is given by
\bea
K_i^\chi(\al) \aeq
\begin{pmatrix} 
-\Phi_i^+ &\Phi_{i}^{-,\chi} \\
\Phi_{i}^{+,\chi}&-\Phi_i^- \\ 
\end{pmatrix}\begin{matrix} 
\dke{00}_i \\
\dke{11}_i \\
\end{matrix} , \la{K_U=0}
\eea
with $\Phi_i^{\pm,\chi}=\sum_b\Phi_{b,i}^{\pm,\chi} $. 
$\{\ke{0}_i {}_i\br{1},\ke{1}_i {}_i\br{0} \}$ part does not relate to the \rd{instantaneous} steady state of $\Hat{K}_i^\chi(\al)$. 
The eigenvalue of the \rd{instantaneous} steady state of $\Hat{K}_i^\chi(\al)$ is given by
\bea
\lm_{i,0}^\chi(\al)=-\f{\Phi_i^+(\al) +\Phi_i^-(\al)}{2} +\sqrt{D_i^\chi(\al)} , \la{lm_i}
\eea
with $D_i^\chi(\al) = [\Phi_i^+ +\Phi_i^-]^2/4-[ \Phi_i^+ \Phi_i^--\Phi_{i}^{-,\chi}\Phi_{i}^{+,\chi}] $. The corresponding left and right eigenvectors are $\dke{\rho_{i,0}^\chi(\al)}=C_i^\chi(\al)\dke{00}_i+E_i^\chi(\al)\dke{11}_i$ and $\dbr{l_{0,i}^\chi(\al)}={}_i\dbr{00}+v_i^\chi(\al){}_i\dbr{11} $ with 
$C_i^\chi(\al)=\f{\Phi_i^{-,\chi}\Phi_i^{+,\chi}}{ [\lm_{i,0}^\chi+\Phi_i^+]^2+\Phi_i^{-,\chi}\Phi_i^{+,\chi}}$, 
$E_i^\chi(\al)=\f{\Phi_i^{+,\chi}(\lm_{i,0}^\chi+\Phi_i^+)}{ [\lm_{i,0}^\chi+\Phi_i^+]^2+\Phi_i^{-,\chi}\Phi_i^{+,\chi}}$, and
\bea
v_i^\chi(\al)=\f{\Phi_i^+-\Phi_i^-+2\sqrt{D_i^\chi(\al)} }{2\Phi_i^{+,\chi}} .
\eea
At $\chi_{b\sig}=0$, $E_i^\chi(\al)$ becomes $E_i(\al)=\f{\Phi_i^+}{\Phi_i^++\Phi_i^-}$ and 
$C_i^\chi(\al)$ becomes $C_i(\al)=1-E_i(\al)$. 

\subsection{\rd{Instantaneous} steady currents}  \la{SC}

The \rd{instantaneous} steady current is given by $I_{b\sig}^{\rm{Steady}}(\al)= \f{\partial \lm_0^\chi(\al) }{\partial (i\chi_{b\sig})} \big \vert_{\chi=0}$. 
In the noninteracting \bl{system}, $\lm_0^\chi(\al)$ is 
$\sum_i \lm_{i,0}^\chi(\al)$ and it leads to $I_{b\sig}^{\rm{Steady}}(\al)=\sum_i I_{i,b\sig}^{\rm{Steady}}(\al)$. 
Here, $I_{i,b\sig}^{\rm{Steady}}(\al)= \f{\partial \lm_{0,i}^\chi(\al) }{\partial (i\chi_{b\sig})} \big \vert_{\chi=0}$ are calculated from \re{lm_i} as
\bea
I_{i,L\sig}^{\rm{Steady}}(\al)=\f{\Ga_{L\sig,i}\Ga_{R,i}(f_R(\tl \om_i)-f_L(\tl \om_i))}{\Ga_i} \la{I_Ls},
\eea
with $\Ga_{b\sig,i}=2 \pi \sum_{k} \abs{W_{bk\sig,i}}^2\dl(\ep_{bk}+\sig g_bB_b-\tl \om_i)$, $\Ga_{b,i}=\sum_\sig \Ga_{b\sig,i}$, and $\Ga_{i}=\sum_b \Ga_{b,i}$.
$I_{i,L\sig}^{\rm{Steady}}(\al)$ vanishes at zero bias ($\be_L=\be_R$, $\mu_L=\mu_R$).
\RD{Let us} consider the modulation of only the thermodynamic parameters ($\al^{\pr\pr}=\{\mu_b,\be_b \}_{b=L,R}$) similar to Refs. [\onlinecite{Main,Utiyama, Watanabe, yoshii2}]. 
The factor depending on $\al^{\pr\pr}$ of $I_{i,b\sig}^{\rm{Steady}}(\al_t)$ is
$(f_{\be_R(t),\mu_R(t)}(\tl \om_i)-f_{\be_L(t),\mu_L(t)}(\tl \om_i))$ with $f_{\be,\mu}(\om)=[e^{\be(\om-\mu)}+1]^{-1}$. Hence
\bea
\bra \Dl N_{b\sig} \ket^{\rm{Steady}}_\tau \aeq \sum_{i}   \f{\Ga_{L\sig,i}\Ga_{R,i}}{\Ga_i} \no\\
&&\hs{-15mm} \times \int_0^\tau dt \ [f_{\be_R(t),\mu_R(t)}(\tl \om_i)-f_{\be_L(t),\mu_L(t)}(\tl \om_i)],
\eea
is generally nonzero and \bl{is much lager} than $\bra \Dl N_{b\sig} \ket^{\rm{Berry}}_S$ because the period $\tau$ is large for adiabatic \bl{pumps}. 
Similarly, we can show that $\bra \Dl N_{b\sig} \ket^{\rm{Steady}}_\tau$ is generally nonzero for interacting system (\res{current,inf}). 
\RD{Reference [\onlinecite{Watanabe}]} considered special modulations of only thermodynamic parameters which satisfy 
$\bra \Dl N_{b\sig} \ket^{\rm{Steady}}_\tau=0$. 
In fact, the \rd{instantaneous} steady currents are always zero for arbitrary modulations of only the dynamics parameters at zero bias. 

\RD{The pumped charge and spin due to the instantaneous steady currents (backgrounds) are generally nonzero even if the time averages of the bias are zero.
References [\onlinecite{Splettstoesser10-2,Splettstoesser12}] chose $V=\mu_L-\mu_R$ as one of the modulating parameters and considered a pumping 
such that $\f{1}{\tau}\int_0^\tau dt \ V(t)=0$ and $\bra \Dl N_{b\sig} \ket^{\rm{Steady}}_\tau \ne 0$. 
In such pumping, the (thermal or voltage) bias is effectively nonzero.}

\RD{Even if the backgrounds do not vanish, 
one can detect the BSN curvatures by subtracting the backgrounds by using zero-frequency measurements or by lock-in measurements. 
However, if one wants to apply the adiabatic pump to the current standard\cite{CS1,CS2}, the instantaneous steady currents should be zero at all times
because the backgrounds are sensitive to the velocity of the modulation of the control parameters and its trajectory. 
In contrast, the pumped charge and spin due to the BSN curvatures are robust against the modulation of the velocity and the trajectory. 
Hence, if one wants to directly apply the BSN curvatures to, for instance, the current standard, one should fix the thermodynamic parameters at zero bias.}

\subsection{BSN curvatures} \la{pump,U=0}

In the following, we consider one level system of which the Hamiltonian is \re{H_S,1} at $U=0$.  
The \rd{instantaneous} steady state is given by $\dke{\rho_0^\chi(\al)}=\otimes_{s=\up,\dw} \dke{\rho_{s,0}^\chi(\al)}$ 
because the Liouvillian is described by a summation ($\Hat{K}^{\chi}= \sum_{s=\up,\dw} \Hat{K}_s^{\chi}$).
Similarly, the corresponding left eigenvalue is given by $\dbr{l_0^\chi(\al)}=\otimes_{s=\up,\dw} \dbr{l_{s,0}^\chi(\al)}$. The BSN vectors \re{BSNv} are given by 
\bea
A_n^{b\sig}(\al) 
\aeq \sum_{s=\up,\dw} v_s^{b\sig}(\al^\pr) \f{\partial E_s(\al) }{\partial \al^n} \la{A_n},
\eea
where  
\bea
v_s^{b\sig}(\al^\pr) \aeq \f{\partial v_s^\chi(\al)}{\partial (i\chi_{b\sig})}  \Big \vert_{\chi=0} 
= \f{\Ga_{b\sig,s}}{\Ga_s} , \la{v_s^bsig}
\eea
with 
\bea
\Ga_{b\sig,s}(\al^\pr) \aeq 2\pi \Dl_b  \sum_{k}\abs{v_{bk\sig,s}}^2 \no\\
&& \times \dl(\ep_{bk}+\sig g_bB_b-\om_0-sg_SB_S). 
\eea
$v_s^{b\sig}(\al^\pr)$ dose not depend on $\al^{\pr\pr}$. Equation (\ref{A_n}) leads to an expression of the BSN curvatures
\bea
F_{mn}^{b\sig}(\al) 
\aeq \sum_{s=\up,\dw} \Big[ \f{\partial v_s^{b\sig}(\al^\pr)  }{\partial \al^m}  \f{\partial E_s(\al) }{\partial \al^n}
-(m \leftrightarrow n) \Big] . \la{F_mn}
\eea

\rd{We emphasize that \re{F_mn} is consistent with the results of Refs. [\onlinecite{Splettstoesser10-2,Splettstoesser12,Main}], which 
showed that the pumped charge (and also spin in \Ref{Splettstoesser12}) vanishes at the noninteracting limit in these settings. 
The set of control parameters $\al$ was $\al^{\pr\pr}$ (for \Ref{Main}) and $\{\om_0, V=\mu_L-\mu_R\}$ (for Refs. [\onlinecite{Splettstoesser10-2,Splettstoesser12}]). 
If $\al^m$ or $\al^n$ is an element of $ \al^{\pr\pr}$, $F_{mn}^{b\sig}(\al)$ is consistently zero.  
In Refs. [\onlinecite{Splettstoesser10-2,Splettstoesser12}], the linewidth functions were energy-independent, namely $\Ga_{b\sig,s}(\al^\pr)=\dl_{\sig,s}\Ga_{b}$=constant. 
Hence $\f{\partial \Ga_{b\sig,s}(\al^\pr) }{\partial \om_0}=0=\f{\partial \Ga_{b\sig,s}(\al^\pr) }{\partial V}$ and $F_{\om_0,V}^{b\sig}(\al)=0$ hold consistently.}

\rd{To calculate $F_{mn}^{b\sig}(\al)$, we need to assume the energy dependences of $\Ga_{b\sig,s}$. 
For the simplicity, we assume that}
\bea
\Ga_{b\sig,s}\aeq  \dl_{\sig,s}[ \Ga_{b}+\Ga_{b}^\pr \cdot(sg_S B_S-\sig g_b B_b) ]\no\\
\aeq \dl_{\sig,s}\Dl_b[ \ga_{b}+\ga_{b}^\pr \cdot(sg_S B_S-\sig g_b B_b) ] , \la{LW}
\eea
where $\Gamma_{b}^\pr$ are energy differential coefficients of linewidth functions at $B_b=B_S=0$. 
Namely, we disregard spin flips induced by tunneling between the \bl{QD} \bl{and the leads}. 
\rd{Equation (\ref{LW}) is always appropriate when $\abs{\Ga_{b}^\pr (g_S B_S-g_bB_b)} \ll \Ga_b $ is satisfied. 
Additionally, we fix} $\al^{\pr\pr}$ to zero bias ($\be_b=\be$, $\mu_b=\mu$), 
\bl{in which $E_s(\al)$ is given by} $E_s(\al)=f(\om_0+sg_SB_S )$ with $f(\om)=[e^{\be(\om-\mu)}+1]^{-1}$.
In this condition, $(\al^m,\al^n)=(B_L,B_S),(\Dl_L,B_S)$ components of the charge and spin BSN curvatures of lead $L$ are 
\begin{widetext}
\bea
F_{B_L,B_S}^{L\up}\pm F_{B_L,B_S}^{L\dw}  \aeq  -g_S g_L \Gamma_{L}^\pr  [ f^\pr(\om_0+g_S B_S)\pm f^\pr(\om_0-g_S B_S)]    \f{\Gamma_R }{ \Ga_\tot^2}\no\\
&&+g_S g_L \Gamma_{L}^\pr  [ f^\pr(\om_0+g_S B_S)\mp f^\pr(\om_0-g_S B_S)] \no\\
&&\hs{3mm} \times \Big( \Gamma_{L}^\pr(g_S B_S- g_L B_L) \f{2\Gamma_R }{ \Ga_\tot^3} +\Gamma_{R}^\pr(g_S B_S- g_R B_R)\f{\Ga_R-\Ga_L}{\Ga_\tot^3} \Big) \la{BL,BS,c,0} ,\\
F_{\Dl_L,B_S}^{L\up} \pm F_{\Dl_L,B_S}^{L\dw} 
\aeq   g_S [f^\pr(\om_0+g_SB_S)\mp f^\pr(\om_0-g_SB_S)]   \f{ \ga_L \ga_R \Dl_R}{ (\ga_L \Dl_L+\ga_R \Dl_R)^2} \no\\
&&+g_S [f^\pr(\om_0+g_SB_S)\pm f^\pr(\om_0-g_SB_S)] \ga_{L}^\pr (g_S B_S- g_LB_L) 
 \f{ \ga_R \Dl_R-  \ga_L \Dl_L}{ (\ga_L \Dl_L+\ga_R \Dl_R)^2} . \la{DL,BS,c,0} 
\eea
\end{widetext}
Here $f^\pr(\om)=\f{\partial f(\om)}{\partial  \om}$ and \bl{$\Ga_\tot=\Ga_L+\Ga_R$}. The pumped charge (spin) induced by a slow cycle modulation of $(\al^n,B_S)$ ($\al^n=B_L, \Dl_L$) are given by
\rd{\bea
&&\hs{-15mm}\bra \Dl N_{L\up} \ket\pm \bra \Dl N_{L\dw} \ket \no\\
\aeq -\int_{S^n} d\al^n dB_S \ (F_{\al^n,B_S}^{L\up}\pm F_{\al^n,B_S}^{L\dw} ), \la{Pump}
\eea
where} $S^n$ are areas enclosed by the trajectories of $(\al^n,B_S)$. \bl{
$F_{\al^n,B_S}^{L\up}\pm F_{\al^n,B_S}^{L\dw}$ ($\al^n=B_L, \Dl_L$) are invariant under the transformation $\ga_b \to c \ga$, $\ga_b^\pr \to c \ga_b^\pr $ (for any $c>0$). 
Hence relevant quantities are $\ga_b^\pr/\Ga_\tot$. The coupling strength $\Ga_\tot$ itself is not important. }
$F_{B_L,B_S}^{L\up}\pm F_{B_L,B_S}^{L\dw}$ are proportional to $g_Sg_L$ and $F_{\Dl_L,B_S}^{L\up} \pm F_{\Dl_L,B_S}^{L\dw}$ are proportional to $g_S$.  
The first terms of the right side of \re{BL,BS,c,0} and \re{DL,BS,c,0} are \RD{dominant} terms. 
\rd{In the limit} $\ga_L^\pr \to 0$, $F_{B_L,B_S}^{L\up}\pm F_{B_L,B_S}^{L\dw}$ and the second term of \re{DL,BS,c,0} vanish; however,  
the \RD{dominant} term of \re{DL,BS,c,0} remai\bl{ns}. At $\om_0=\mu$, $f^\pr(\om_0+g_SB_S)- f^\pr(\om_0-g_SB_S)$ vanish. 
Hence, at $\om_0=\mu$, the \RD{dominant} terms of the spin BSN curvature of $(B_L,B_S)$ pump and 
the charge BSN of $(\Dl_L,B_S)$ pump vanish. 
The contour plots of these BSN curvatures are shown in Figs. \ref{Bs,Bl}(a) and 1(b) and Figs. \ref{Bs,Dl}(a) and 2(b). The details are explained in \res{pump}.

It is important to remark that $(\al^m,\al^n)=(B_L,B_R),(\Dl_L,\Dl_R)$ components of the charge and spin BSN curvatures are zero at zero bias 
because, in \re{F_mn}, $E_s(\al)=f(\om_0+sg_SB_S)$ are independent of $B_{L/R}$ and $\Dl_{L/R}$.

\section{Interacting system} \la{Interacting}

In this section, we study the interacting system \re{H_S,1}.
First, we explain the Liouvillian for $0 \le U \le \infty$ (\res{Liouvillian}). Next, the \rd{instantaneous} steady charge and spin currents are calculated at $U=\infty$ (\res{current,inf}). 
\rd{In \res{pump}, we confirm the consistency between our results and \Ref{Splettstoesser12} for $0 \le U \le \infty$}.  
Finally, the BSN curvatures corresponding to \re{BL,BS,c,0} and \re{DL,BS,c,0} are calculated at $U=\infty$ and differences of the results 
between $U=0$ and $U=\infty$ are discussed. 

\subsection{Liouvillian} \la{Liouvillian}

\rd{We explain the Liouvillian for $k_{\rm{B}}T > \Ga$, in which the Born-Markov approximation is appropriate.} 
The matrix representation of the Liouvillian of RWA by the number states $\{\ke{n_\up n_\dw} \}$ ($n_s=0,1$ are the numbers of an electron with spin $s=\up,\dw$) 
is a $16 \times 16$ matrix which is block diagonalized
 to the ``diagonal"  space (spanned by  \rd{$\{\ke{n_\up n_\dw}\br{n_\up n_\dw}\}_{n_\up,n_\dw=0,1}$}) and the ``off-diagonal" space 
(spanned by \rd{$\{\ke{n_\up n_\dw}\br{m_\up m_\dw} \}_{(n_\up,n_\dw)\ne (m_\up,m_\dw)}$}). The ``diagonal" block is given by
\begin{widetext}
\bea
K^\chi(\al) \aeq  \begin{pmatrix} 
-[\Phi_{\up}^++\Phi_{\dw}^+] &\Phi_{\up}^{-,\chi}&\Phi_{\dw}^{-,\chi}&0\\
\Phi_{\up}^{+,\chi} &-[\Phi_{\up}^-+\phi_{\dw}^+]&0&\phi_{\dw}^{-,\chi}\\ 
\Phi_{\dw}^{+,\chi} &0&-[\Phi_{\dw}^-+\phi_{\up}^+]&\phi_{\up}^{-,\chi}\\ 
0 &\phi_\dw^{+,\chi}&\phi_\up^{+,\chi}&-[\phi_{\up}^-+\phi_{\dw}^-]  \\ 
\end{pmatrix}\begin{matrix} 
\dke{0000} \\
\dke{1010} \\
\dke{0101} \\
\dke{1111} \\
\end{matrix} \hs{2mm},  \la{K_int}
\eea
\end{widetext} 
with 
\bea
\phi_{b,s}^{\pm,\chi} \aeq 2\pi \Dl_b  \sum_{k,\sig}\abs{v_{bk\sig,s}}^2 f_b^\pm(\om_0+sg_SB_S+U) \no\\
&&\hs{-3mm}\times e^{\mp i\chi_{b\sig}} \dl(\ep_{bk}+\sig g_bB_b-\om_0-sg_SB_S-U), 
\eea
and $\Phi_{b,s}^{\pm,\chi}=\phi_{b,s}^{\pm,\chi} \vert_{U=0}$. The off-diagonal block is a $(12\times12)$-diagonal matrix, which dose not relate to the \rd{instantaneous} steady state.
At $U=0$, $K^\chi(\al)$ becomes $K_\up^\chi(\al)\otimes 1_\dw+1_\up \otimes K_\dw^\chi(\al)$, where $K_s^\chi(\al)(s=\up,\dw)$ are given by \re{K_U=0} and $1_s$ are identity matrices. 
In the opposite limit $U \to \infty$, $K^\chi(\al)$ reduces to
\bea
K^{\chi(\infty)}(\al)
\aeq  \begin{pmatrix} 
-[\Phi_{\up}^++\Phi_{\dw}^+] &\Phi_{\up}^{-,\chi}&\Phi_{\dw}^{-,\chi}\\
\Phi_{\up}^{+,\chi} &-\Phi_{\up}^-&0\\ 
\Phi_{\dw}^{+,\chi} &0&-\Phi_{\dw}^-\\ 
\end{pmatrix} \begin{matrix} 
\dke{0000} \\
\dke{1010} \\
\dke{0101} \\
\end{matrix} , \la{K_int_inf}
\eea
because the density of state of both leads vanishes at high energy ($\phi_{s}^\pm \to 0).$

\begin{figure*}
\includegraphics[width=16cm]{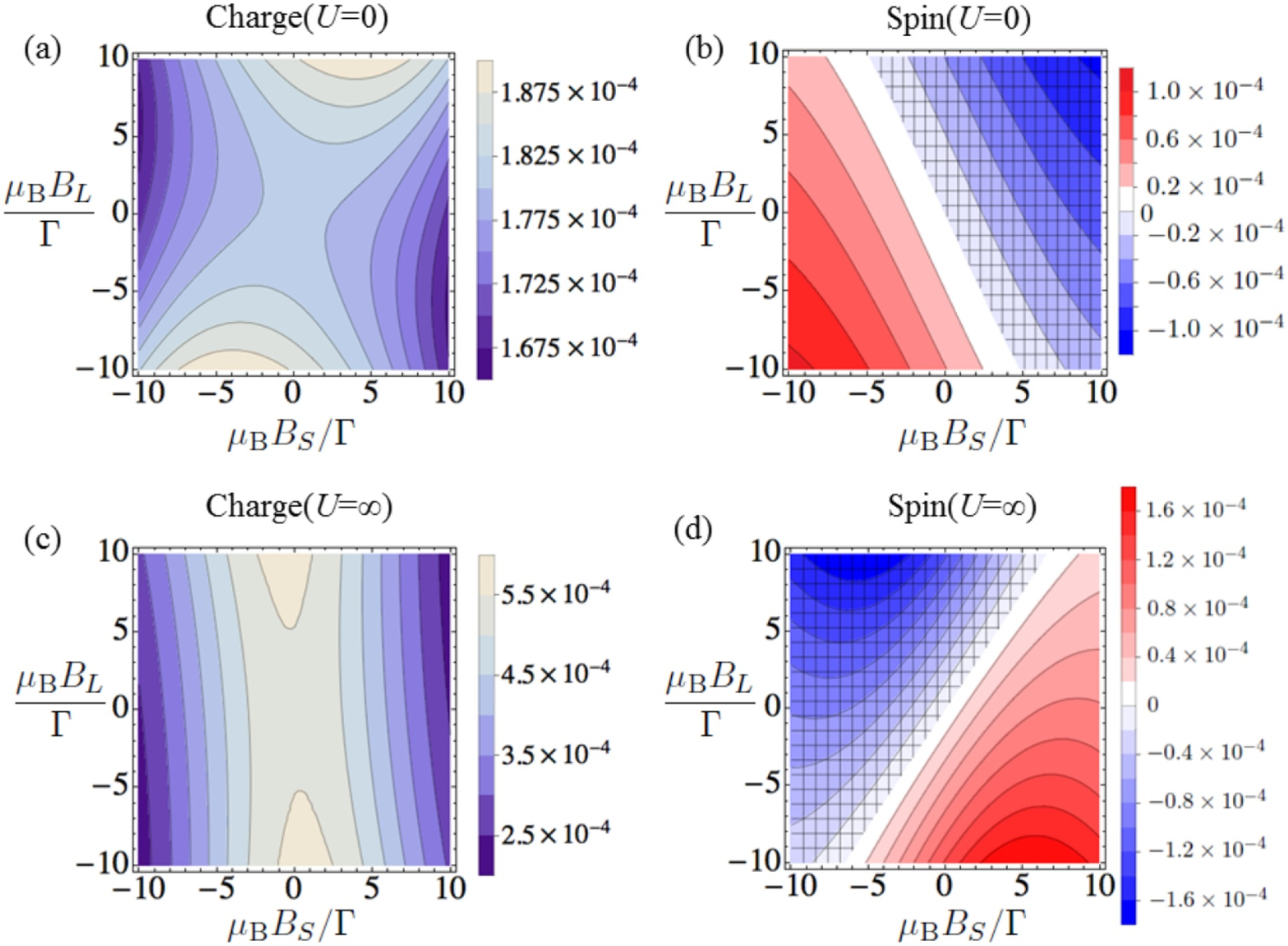}
\caption{\label{Bs,Bl}(Color online)(a) BSN curvature of charge of $(B_L,B_S)$ pump, \bl{$[F_{B_L,B_S}^{L\up}+\RD{F_{B_L,B_S}^{L\dw}}]/\big(\f{\mu_{\rm{B}}}{\Ga}\big)^2$} at $U=0$, 
(b) the BSN curvature of spin, \bl{$[F_{B_L,B_S}^{L\up}-\RD{F_{B_L,B_S}^{L\dw}}]/\big(\f{\mu_{\rm{B}}}{\Ga}\big)^2$} at $U=0$, 
(c) \bl{$[F_{B_L,B_S}^{L\up}+\RD{F_{B_L,B_S}^{L\dw}}]/\big(\f{\mu_{\rm{B}}}{\Ga}\big)^2$} at $U=\infty$, and
(d) \bl{$[F_{B_L,B_S}^{L\up}-\RD{F_{B_L,B_S}^{L\dw}}]/\big(\f{\mu_{\rm{B}}}{\Ga}\big)^2$} at $U=\infty$. 
The values of the parameters used for these plots are $\Ga_L=\Ga_R=\Ga$, $\Ga_L^\pr=\Ga_R^\pr=0.1$, $\beta=0.5/\Ga$, $\om_0=\mu-3\Ga$, and $B_R=0$, 
and all $g$ factors ($g_L^\ast$, $g_R^\ast$, $g_S^\ast$) are $-0.44$ (bulk GaAs). \rd{The hatched areas of (b),(d) denote negative value.}}
\end{figure*}
\begin{figure*}
\includegraphics[width=16cm]{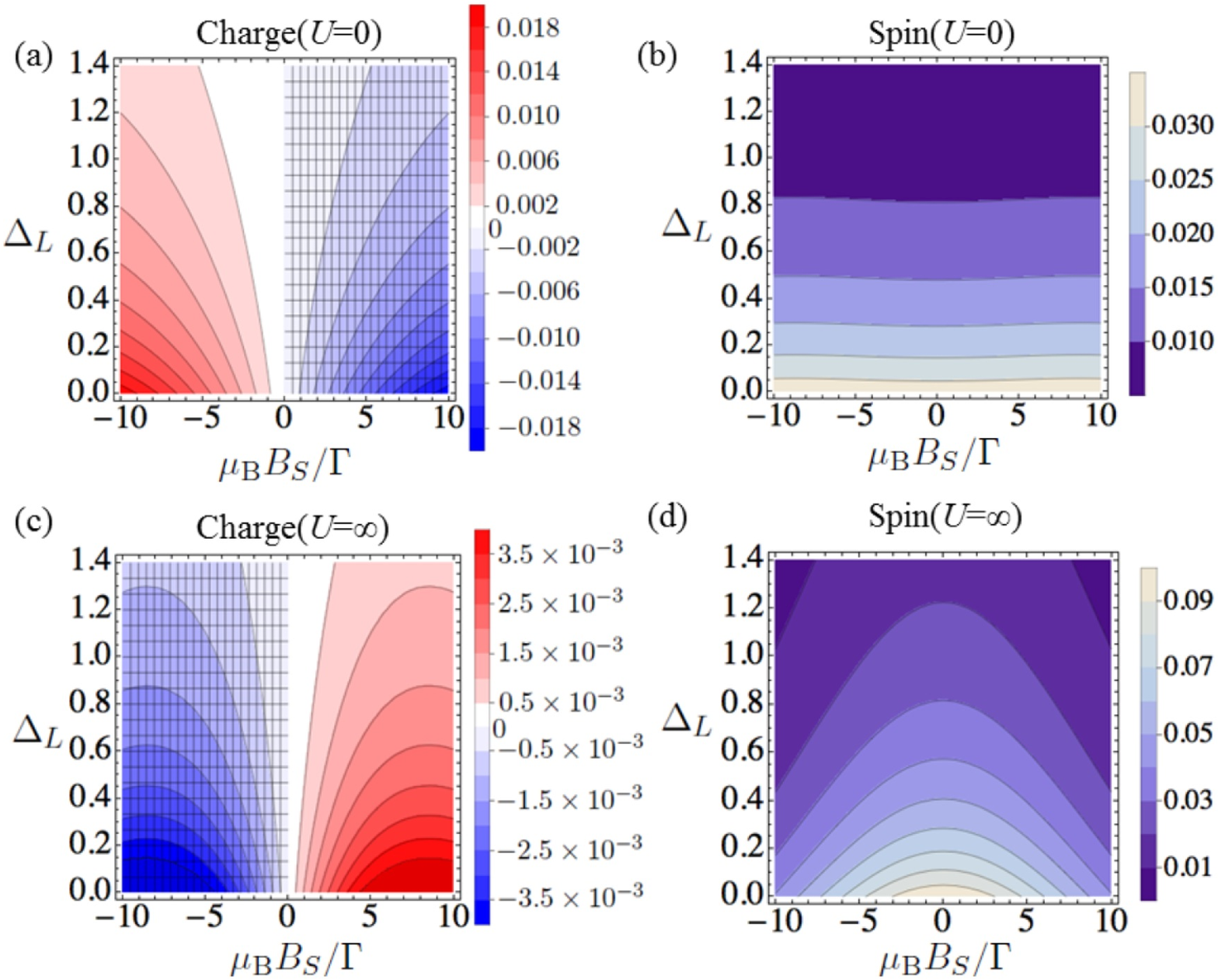}
\caption{\label{Bs,Dl}(Color online)(a) BSN curvature of charge of $(\Dl_L,B_S)$ pump, \bl{$[F_{\Dl_L,B_S}^{L\up}+\RD{F_{\Dl_L,B_S}^{L\dw}}]/\f{\mu_{\rm{B}}}{\Ga}$} at $U=0$, 
(b) the BSN curvature of spin, \bl{$[F_{\Dl_L,B_S}^{L\up}-\RD{F_{\Dl_L,B_S}^{L\dw}}]/\f{\mu_{\rm{B}}}{\Ga}$} at $U=0$, 
(c) \bl{$[F_{\Dl_L,B_S}^{L\up}+\RD{F_{\Dl_L,B_S}^{L\dw}}]/\f{\mu_{\rm{B}}}{\Ga}$} at $U=\infty$, and 
(d) \bl{$[F_{\Dl_L,B_S}^{L\up}-\RD{F_{\Dl_L,B_S}^{L\dw}]}/\f{\mu_{\rm{B}}}{\Ga}$} at $U=\infty$. 
The values of the parameters used for these plots are $\ga_L=\Ga_R=\Ga$, $\ga_L^\pr=\Ga_R^\pr=0.1$\bl{, and $B_L=0$} and other conditions are \bl{the} same as Fig. \ref{Bs,Bl}. 
\rd{The hatched areas of (a),(c) denote negative value.}}
\end{figure*}
\begin{figure*}
\includegraphics[width=16cm]{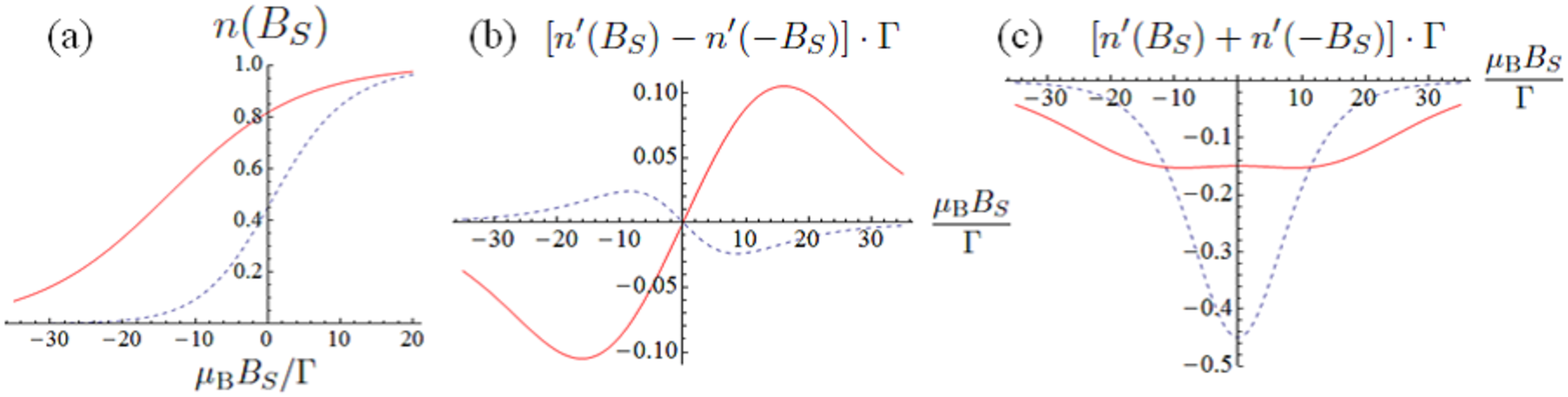}
\caption{\label{fign+-}(Color online) (a) \rd{$n(B_S)=f(\om_0+g_SB_S)$} (solid line) \RD{or} $\rho(B_S)$ (dashed line). (b) \rd{$n^\pr(B_S)-n^\pr(-B_S)$, and \rd{(c)} $n^\pr(B_S)+n^\pr(-B_S)$}, 
where $n^\pr(\pm B_S) =\f{1}{g_S}\f{\partial n(B)}{\partial B} \vert_{B=\pm B_S}$. 
In all plots, $\beta=0.5/\Ga$, $\om_0=\mu-3\Ga$, and \RD{$g_S=-0.44 \times \mu_{\rm{B}}/2$.}}
\end{figure*}

\subsection{\rd{Instantaneous} steady currents} \la{current,inf}

In this section, we set $U=\infty$. The characteristic polynomial of $K^{\chi(\infty)}$ is denoted as
$C_3(\chi,\lm)=\det(K^{\chi(\infty)}-\lm )=\sum_{n=0}^2 C_n(\chi)\lm^n-\lm^3$. 
Because of $C_0(0)=0$, $\lm=0$ is one of the solutions at $\chi=0$.
Now we set $\chi_{b\sig}$ as infinitesimal and other counting fields are zero. 
Then, the eigenvalue corresponding to the \rd{instantaneous} steady state is given by 
$\lm=\lm_0(\chi)=i \chi_{b\sig} \cdot I_{b\sig}^{\rm{Steady}}+\mathcal{O}(\chi_{b\sig}^2)$. 
It leads to $0=C_3(\chi,\lm_0(\chi))=C_1(0)i\chi_{b\sig} I_{b\sig}^{\rm{Steady}} + i\chi_{b\sig} C_0^{b\sig}$ with
$C_0^{b\sig}=\f{\partial C_0(\chi)}{\partial (i\chi_{b\sig})} \big \vert _{\chi=0}$, and we obtain
\bea
I_{b\sig}^{\rm{Steady}} \aeq -\f{C_0^{b\sig}}{C_1(0)} , \la{key_tec}
\eea
with $C_1(0) = -[\Phi_{\up}^+\Phi_{\dw}^-+\Phi_{\up}^-\Phi_{\dw}^++\Phi_{\up}^-\Phi_{\dw}^-]$. From $C_0(\chi) = -[\Phi_{\up}^++\Phi_{\dw}^+]\Phi_{\up}^-\Phi_{\dw}^- 
+\Phi_{\dw}^{-,\chi}\Phi_{\up}^-\Phi_{\dw}^{+,\chi} +\Phi_{\dw}^-\Phi_{\up}^{-,\chi}\Phi_{\up}^{+,\chi}$, we have
\bea
I_{L\sig}^{\rm{Steady}} (\al)\aeq \f{\sum_{s=\up,\dw}  \Phi_{-s}^-\Gamma_{L\sig,s}\Gamma_{R,s} (f_{R,s}-f_{L,s} ) }
{\Phi_{\up}^+\Phi_{\dw}^-+\Phi_{\up}^-\Phi_{\dw}^++\Phi_{\up}^-\Phi_{\dw}^-} ,
\eea
\rd{where $\Phi_{-s}^-$ $(s=\up,\dw)$ describes $\Phi_\dw^-$ for $s=\up$ and $\Phi_\dw^-$ for $s=\dw$}. At zero bias, the \rd{instantaneous} steady currents vanish. Similar to \res{SC}, 
$\bra \Dl N_{b\sig} \ket^{\rm{Steady}}_\tau$ are generally nonzero when $\al^{\pr\pr}$ is not fixed at zero bias.

\subsection{BSN curvatures} \la{pump}

The \rd{instantaneous} steady state $\rho_0(\al) $ and corresponding left eigenvector $l_0^\chi(\al)$ are written as 
$\rho_0 = \rho_{0}\ke{00} \br{00} +\rho_{\up}\ke{10} \br{10} +\rho_{\dw}\ke{01} \br{01} +\rho_{2}\ke{11} \br{11}$ and
$l_0^\chi=  \ke{00} \br{00} +l_{\up}^\chi \ke{10} \br{10} +l_{\dw}^\chi \ke{01} \br{01} +l_2^\chi \ke{11} \br{11}$. The BSN vectors are given by
\bea
A_n^{b\sig}(\al) 
\aeq \sum_{c=\up,\dw,2}  l_c^{b\sig}(\al) \f{\partial \rho_c(\al) }{\partial \al^n} ,
\eea
where $l_c^{b\sig}(\al) = \f{\partial [l_c^{\chi}(\al)]^\ast} {\partial (i\chi_{b\sig})} \big \vert_{\chi=0}  $. It leads to the BSN curvatures
\bea
F_{mn}^{b\sig}(\al) 
\aeq \sum_{c=\up,\dw,2}  \f{\partial l_c^{b\sig}(\al) }{\partial \al^m}   \f{\partial \rho_c(\al) }{\partial \al^n}  -(m \leftrightarrow n) . \la{F_mn_U_fin}
\eea
\rd{We confirmed the consistency between our results and \Ref{Splettstoesser12}, which studied the similar system for $0\le U \le\infty$ using \BL{the wideband limit}. 
As we explained at \res{Nona}, $\varphi_\ka^{[\mu]}(\al)$ of \re{varphi} corresponds to $-\dbr{l_0^\mu(\al)}$, namely $-l_c^{b\sig}(\al)$. 
In the condition of \BL{the wideband limit [i.e., \re{LW} with $\Ga_b^\pr=0$]}, we calculated $l_c^{b\sig}(\al)$ $(c=\up,\dw,2)$ for $0\le U \le \infty$ and
confirmed numerically the correspondence between $\varphi_c^{[\mu]}(\al)$ ($c=\up,\dw,2$) and $-[l_c^{b\up}(\al)\pm l_c^{b\dw}(\al) ]$ for the charge and spin pump.}

Particularly, \rd{in the limit} $U \to \infty$, $\rho_{2}$ vanishes and $F_{mn}^{b\sig}(\al) $ reduces to
\bea
F_{mn}^{b\sig(\infty)}(\al) 
\aeq \sum_{s=\up,\dw} \f{\partial l_s^{b\sig(\infty)}(\al) }{\partial \al^m}   \f{\partial \rho_s^{(\infty)}(\al) }{\partial \al^n} -(m \leftrightarrow n) , \la{F_mn_U_inf}
\eea
where $\rho_s^{(\infty)}(\al)$ and $l_s^{b\sig(\infty)}(\al)$ are the \rd{limits} $U \to \infty$ of $\rho_s(\al)$ and $l_s^{b\sig}(\al)$, respectively. From \re{K_int_inf} we obtain
\bea
\rho_s^{(\infty)}(\al) \aeq \f{ \Phi_s^+\Phi_{-s}^-}{\Phi_{\up}^- \Phi_{\dw}^-+\Phi_{\up}^-\Phi_{\dw}^++\Phi_{\up}^+\Phi_{\dw}^-} ,\\
\{l_s^{(\infty)}(\al)\}^\ast \aeq \f{\Phi_s^{-,\chi}}{\Phi_s^- +\lm_0^\chi} ,
\eea
and
\bl{\bea
l_s^{b\sig(\infty)}(\al)
\aeq \f{\Ga_{b\sig,s}(1-f_b(\om_s))-I_{b\sig}^{\rm{Steady}}(\al)}{\Phi_s^-} ,
\eea
w}ith $\om_s=\om_0+\RD{sg_SB_S}$. In the following, we fix $\al^{\pr\pr}$ to zero bias ($\be_b=\be$, $\mu_b=\mu$) \bl{and suppose \re{LW}}. 
Then, $l_s^{b\sig(\infty)}(\al)$ equals $v_s^{b\sig}(\al^\pr)$ given by \re{v_s^bsig} and  $\rho_s^{(\infty)}(\al)$ are \bl{given by}
\bea
\rho(sB_S) \aeq \f{ e^{-\be(\om_s-\mu)}}{1+ e^{-\be(\om_\dw-\mu)}+e^{-\be(\om_\up-\mu)} }.
\eea
\rd{We emphasize that} \bl{$F_{mn}^{b\sig(\infty)}(\al)$} can be obtained by \rd{just} a replacement,
\bea
E_s(\al)=f(\om_s) \to \rho(sB_S),
\eea
in \re{F_mn}. The charge and spin BSN curvatures of $(B_L,B_S),(\Dl_L,B_S)$ pump are given by a replacement $f^\pr(\om_0\pm g_S B_S) \to \rho^\pr(\pm B_S)$ in Eqs. (\ref{BL,BS,c,0}) 
and (\ref{DL,BS,c,0}), where
$\rho^\pr(B_S) \defe \f{1}{g_S }\f{\partial \rho(B_S)}{\partial B_S}  $. 
Similar to $U=0$, the charge and spin BSN curvatures of $(B_L,B_R),(\Dl_L,\Dl_R)$ pump are zero.

In Figs. \ref{Bs,Bl}(a)-1(d), we plot the BSN curvatures of $(B_L,B_S)$ pump \bl{normalized by $(\mu_{\rm{B}}/\Ga)^2$, where $\Ga=\Ga_L=\Ga_R$ 
and $\mu_{\rm{B}}=57.88$ $\mu$eV/T is the Bohr magneton.}
For $U=0$, the charge and spin BSN curvatures  are shown in Fig. \ref{Bs,Bl}(a) and Fig. \ref{Bs,Bl}(b), and for $U=\infty$ these are shown in Figs. \ref{Bs,Bl}(c) and 1(d).
The vertical and horizontal axes of these plots are the strength of magnetic fields $B_S$, $B_L$ normalized by $\Ga/\mu_{\rm{B}}$.
The values of the parameters used for these plots are $\Ga_L=\Ga_R=\Ga$, $\Ga_L^\pr=\Ga_R^\pr=0.1$, $\beta=0.5/\Ga$, $\om_0=\mu-3\Ga$, $B_R=0$, 
and $g_L^\ast=g_R^\ast=g_S^\ast=-0.44$ (bulk GaAs).
The BSN curvatures of $(\Dl_L,B_S)$ pump \bl{normalized by $\mu_{\rm{B}}/\Ga$} are shown similarly in Figs. \ref{Bs,Dl}(a)-2(d). 
In all plots, $\ga_L=\Ga_R=\Ga$, $\ga_L^\pr=\Ga_R^\pr=0.1$\bl{, $B_L=0$}, and other conditions are the same as in Fig. \ref{Bs,Bl}. 
\rd{In Figs. \ref{Bs,Bl} and \ref{Bs,Dl}, the maximum values of $\abs{\Ga_{b}^\pr (g_S B_S-g_bB_b)}/\Ga_b$ are 0.44 and 0.22 ($<$1), respectively.}
The pumped charges and spins are given by \re{Pump}. 

Figure \ref{fign+-}(a) shows the \RD{instantaneous} average numbe\bl{rs} of the up spin electr\bl{on} of the \bl{QD,} $n(B_S)=f(\om_0+g_SB_S)$ (for $U=0$, solid line) 
\RD{or} $\rho(B_S)$ (for $U=\infty$, dashed line) for 
$\beta=0.5/\Ga$, $\om_0=\mu-3\Ga$, and $g_S=-0.44 \times \mu_{\rm{B}}/2$.
Because two electrons cannot occupy a \bl{QD} at $U=\infty$, the magnetic \bl{field} dependence of $\rho(B_S)$ is more sensitive than $f(\om_0+g_SB_S)$.
Figures \ref{fign+-}(b) and 3(c) show $n^\pr(B_S)\mp n^\pr(-B_S)$ normalized by \bl{$1/\Ga$}, where $n^\pr(\pm B_S) =\f{1}{g_S}\f{\partial n(B)}{\partial B} \vert_{B=\pm B_S}$.

\rd{In Figs. \ref{Bs,Dl}(a) and 2(c), the charge BSN curvatures of $(\Dl_L,B_S)$ pump vanish at $B_S=0$. 
This is because the first term of \re{DL,BS,c,0} vanishes since $n^\pr(B_S)-n^\pr(-B_S)=0$ for $B_S=0$ and 
the second term vanishes since $g_SB_S-g_bB_b=0$ for $B_S=0=B_L$. Similarly, in Figs. \ref{Bs,Bl}(b) and (d), the spin BSN curvatures of $(B_L,B_S)$ pump vanish at $B_S=0=B_L$. 
The zero lines in these plots relate to the cancellation between the first and second terms of \re{BL,BS,c,0}. 
Figures. \ref{Bs,Bl}(a),1(c) and Figs. \ref{Bs,Bl}(b),1(d) are respectively symmetric and antisymmetric under the transformation $(B_S,B_L) \to (-B_S,-B_L)$. 
Similarly, Figs. \ref{Bs,Dl}(b),2(d) and Figs. \ref{Bs,Dl}(a),2(c) are respectively symmetric and antisymmetric under the transformation $B_S \to -B_S$. 
We emphasize that pure charge and pure spin pumps are respectively realized for $(B_L,B_S)$ pump and $(\Dl_L,B_S)$ pump such that the areas $S^n$ in \re{Pump} are symmetric under the above transformations. 
An instance of symmetric area of $(B_L,B_S)$ pump is a \RD{disk} of which the center is $B_S=0=B_L$.}
 
In $\om_0>\mu$ region, the larger $\om_0-\mu$, the less \BL{difference} between $U=0$ and $U=\infty$ becomes.
The Coulomb interaction prevents two electrons from occupying the \bl{QD}. 
This effect is conspicuous in the $\om_0<\mu$ region, although it is not important in the $\om_0>\mu$ region.

As shown in Figs. \ref{Bs,Bl}(a),1(c) and Figs. \ref{Bs,Dl}(b),2(d), the $B_S$ dependence of the charge BSN curvature of $(B_L,B_S)$ pump 
and the spin BSN curvature of $(\Dl_L,B_S)$ pump at $U=0$
are more gentle than those at $U=\infty$. It results from the behavior of $n^\pr(B_S)+n^\pr(-B_S)$ as shown in Fig. \ref{fign+-}(c).

As shown in Figs. \ref{Bs,Bl}(b),1(d) and in Figs. \ref{Bs,Dl}(a),2(c), the $B_S$ dependence of the spin BSN curvature of $(B_L,B_S)$ pump 
and the charge BSN curvature of $(\Dl_L,B_S)$ pump 
are opposite. This is because the leading term (in weak magnetic field region) of these are proportional to $n^\pr(B_S)- n^\pr(-B_S)$ and 
its $B_S$ dependence is opposite in $U=0$ and $U=\infty$ for $\om_0-\mu<0$ as indicated in Fig. \ref{fign+-}(b).
This inversion is realized for only $\om_0-\mu<0$ region. At $\om_0=\mu$, $f^\pr(\om_0+g_SB_S)-f^\pr(\om_0-g_SB_S)$ vanish. 
In $\om_0>\mu$ region, the signs of $f^\pr(\om_0+g_SB_S)-f^\pr(\om_0-g_SB_S)$ and $\rho^\pr(B_S)-\rho^\pr(-B_S)$ are the same. 

\rd{In Figs. \ref{Bs,Bl} and \ref{Bs,Dl}, absolute values of the normalized BSN curvatures are smaller than unity. However, we can improve this problem by tuning $g$ factors. }
The first and second terms of the right side of \re{BL,BS,c,0} are the second and third \rd{order in the $g$ factors}, 
and the first and second terms of the right side of \re{DL,BS,c,0} are the first and second \rd{order in the $g$ factors}.
If all $g$ factors change to $-20$ (for example for the materials like InAs, InSb), the first, second, \bl{and} third order terms 
become about $45$, $2\ 000$, \bl{and} $90\ 000$ times.
\rd{In fact, for these values of $g$ factors, the assumption \re{LW} is not appropriate for magnetic fields that are not small; we need concrete energy dependence of linewidth functions.}

\section{Conclusion} \la{Con}

In this paper, we investigated quantum adiabatic pump of charge and spin using the \rd{FCS-QME (full counting statistics with quantum master equation)} approach 
proposed by Yuge {\it et al.}\cite{Main}.
\rd{We} studied the nonadiabatic effect and showed the correspondence between our approach and the real-time diagrammatic approach \cite{Splettstoesser12}(\res{Nona}), 
\rd{and confirmed the consistency between the two methods in the concrete model, the one level interacting quantum dot \bl{(QD)} (\res{pump,U=0} and \res{pump})}. 
Moreover, \rd{in \res{Nona}}, we showed that the Berry-Sinitsyn-Nemenman (BSN) phase derived under the \rd{``adiabatic" condition (which makes the Berry phase like treatment appropriate)} 
has the nonadiabatic information. 
The \rd{FCS-QME} picks out one higher order nonadiabatic piece of information from the solution of the QME. 
Particularly, the \rd{instantaneous} steady state (the zeroth order of the pumping frequency) gives first order response (pumped current).
This fact may be related to \Ref{yuge2} which is connected the BSN vector and the liner response theory of the QME. 

We generalized the \rd{FCS-QME} approach to the multicounting field (\res{FCS-QMEd}) and studied the \bl{QDs} system weakly coupled to leads ($L$ and $R$) in \res{Non-interacting} 
and \res{Interacting}. 
We showed that the pumped charge and spin coming from the \rd{instantaneous} steady current are not negligible when the thermodynamic parameters (the chemical potentials 
and the temperatures of leads) are not fixed to zero bias
in noninteracting \bl{QDs} (\res{SC}) and an interacting \bl{QD} (\res{current,inf}). 
\rd{To observe the spin effects, we consider collinear magnetic fields, which relate to spins through the Zeeman effect, with different amplitudes applying to the \bl{QDs} ($B_S$) and the leads ($B_L$ and $B_R$). 
We focused on the dynamic parameters ($B_S$, $B_{L/R}$} and the coupling strength between \bl{QDs} and leads, $\Dl_{L/R}$) as control parameters. 

In one level \bl{QD} with the Coulomb \BL{interaction} $U$, we analytically calculated the BSN curvatures of spin and charge of $(B_L,B_S)$ pump 
and $(\Dl_L,B_S)$ pump for the noninteracting limit ($U=0$) and the strong interaction limit ($U=\infty$)
using the rotating wave approximation (RWA) defined as the long coarse-graining time limit of the coarse-graining approximation (CGA).
The difference between $U=0$ and $U=\infty$ appeared \rd{through the \RD{instantaneous} averages of the numbers of the electron with up and down spin in \bl{QD}}. 
\rd{For $(B_L, B_S)$ pump, the energy dependences of linewidth functions, which are usually neglected, are essential}.
Additionally, the adiabatic modulations of $(B_L,B_R)$ or $(\Dl_L,\Dl_R)$ can pump \rd{neither charge nor spin}.

In this paper, only $U=0$ and $U=\infty$ limits are studied. \bl{In fact}, we can analyze finite $U$ based on \re{F_mn_U_fin}. This is our future work.
Recently, Yoshii and Hayakawa \cite{yoshii2} studied adiabatic pump of charge by only the thermodynamic parameters using the same approach 
in a similar system without magnetic fields (for finite $U$).
The work (for the thermodynamic parameters) involving the above problem is contrastive to our work (for the dynamic parameters).

\acknowledgments

We acknowledge helpful discussions with S. Ajisaka, K. Watanabe, R. Yoshii, H. Hayakawa, and N. Taniguchi 
\bl{Particularly, we acknowledge Yu Watanabe for discussions about the derivation of \re{W_R}.} 
Part of this work is supported by JSPS \bl{KAKENHI (26247051).}

\appendix

\section{Liouville space} \la{Liouville space}

By following correspondence, an arbitrary linear operator (which operates to the Hilbert space) $\bu =\sum_{n,m} \br{n}\bu \ke{m} \ke{n}\br{m}$ is mapped to a vector of the 
Liouville space\cite{Fano,FCS-QME}, $\dke{\bu}=\sum_{n,m} \br{n}\bu \ke{m} \dke{nm} $:
\bea
 \ke{n}\br{m} \!\!\! &\taiou& \!\!\! \dke{nm} ,\la{taiou1} \\
 \tr(\ke{m}\br{n} n^\pr \ket \br{m^\pr} )  \!\!\! &\taiou& \!\!\! \dbra nm \vert n^\pr m^\pr \dket , \\
 \tr(A\dg B) \!\!\! &\taiou& \!\!\! \dbra A \vert B \dket ,\la{taiou3}  \\
\tr(\bu)  \!\!\! &\taiou& \!\!\! \dbra 1\dke{\bu} .
\eea
Here, $\{\ke{n}\}$ is an arbitrarily complete orthonormal basis. The inner product of the Liouville space is defined by the Hilbert-Schmidt product [\re{taiou3}]. The Hermitian conjugate of $\dke{\bu}$ is defined as $\dbr{\bu}=(\dke{\bu})\dg=\sum_{n,m} \br{n}\bu \ke{m}^\ast \dbr{nm} $.
An arbitrary linear superoperator $\Hat{J}$ which operates to any operator ($\bu$) is mapped to a corresponding operator of the Liouville space ($\check{J}$) as
\bea
\dke{\Hat{J}\bu} \aeq \check{J}\dke{\bu} .
\eea
The matrix representation of $\check{J}$ (or $\Hat{J}$) is defined by 
\bea
J_{nm,kl} \aeq \dbr{nm} \check{J} \dke{kl} .
\eea
In the main text of this paper, both $\check{J}$ and $\Hat{J}$ are denoted by $\Hat{J}$.

Generally, the Liouvillian \rd{$\Hat{K}^\chi$} operates to an operator $\bu$ as
\rd{\bea
\Hat{K}^\chi\bu \aeq -i[H_S,\bu]+\Hat{\Pi}^\chi \bu ,\la{K}\\
\Hat{\Pi}^\chi \bu \aeq \sum_a A_a^\chi \bu B_a ,
\eea
where $H_S$ is the system Hamiltonian, $\Hat{\Pi}^\chi$ is the dissipator like \re{CG_go} and $A_a^\chi$, $B_a$ are operators. 
The matrix representation of \re{K} is given by
\bea
\sum_{k,l} K_{nm,kl}^\chi \bu_{kl}
\aeq \sum_{k,l} \big[-i\{ (H_S)_{nk} \dl_{lm} -\dl_{nk} (H_S)_{lm} \} \bu_{kl} \no\\
&&+\{ \sum_a (A_a^\chi)_{nk} (B_a)_{lm} \} \bu_{kl} \big] ,
\eea
where $\bu_{kl}=\br{k}\bu \ke{l}$.  Hence the matrix representation of $\Hat{K}^\chi$ is given by}\rd{
\bea
K_{nm,kl}^\chi \aeq -i H_{nm,kl}+\Pi_{nm,kl}^\chi , \la{mat_rp1}\\
H_{nm,kl} \aeq (H_S)_{nk} \dl_{lm} -\dl_{nk} (H_S)_{lm} ,\la{mat_rp2} \\
\Pi_{nm,kl}^\chi  \aeq  \sum_a (A_a^\chi)_{nk} (B_a)_{lm} \la{mat_rp3} .
\eea
}

\rd{Finally, we consider the current operators defined by \re{defW}. $\Hat{K}^\mu=\f{\partial \Hat{K}^\chi(\al) }{\partial (i\chi_\mu)} \big \vert_{\chi=0}$ is given by
\bea
\Hat{K}^\mu \bu \aeq \sum_a A_a^\mu \bu B_a.
\eea
Hence the current operators defined by \re{defW} are given by
\bea
W_\mu \aeq \sum_a B_a A_a^\mu. \la{W_AB}
\eea}

\section{Details of FCS-QME} \la{CGA}

$\rho_\tot^\chi(t)$(\res{FCS-QMEd}) is governed by the \rd{modified} von Neumann equation\cite{FCS-QME}: 
\bea 
\f{d}{dt} \rho_\tot^\chi(t) \aeq  -i[H_\tot(t), \rho_\tot^\chi(t)]_\chi , \la{GLN}
\eea
where $H_\tot(t)$ is the total Hamiltonian and $[A,B]_\chi = A_{\chi}B-BA_{-\chi}$ with $A_\chi=e^{i \sum_\mu \chi_\mu O_\mu/2}Ae^{-i\sum_\mu \chi_\mu O_\mu/2} $. 
$H_\tot(t)$ is given by \rd{$H_\tot(t)=H_S(t)+\sum_{b}[H_b(t)+H_{Sb}(t)] $}, where $H_S$ is the system (denoted by $S$) Hamiltonian, $H_b$ is the Hamiltonian of 
bath $b$, and \rd{$H_{Sb}$} is the tunneling interaction Hamiltonian between $S$ and bath $b$. 
In the following, we suppose \bl{Eqs. (\ref{H_b}), (\ref{H_S}) and (\ref{H_1}}) for an arbitrary number of leads ($b=1,2,\cdots,M$). 
\rd{The initial condition of $\rho_\tot^\chi(t)$ is given by \cite{FCS-QME}
$ \rho_\tot^\chi(0)=\sum_{\{ o_\nu \}} P_{\{ o_\nu \}} \rho_\tot(0) P_{\{ o_\nu \}}$. 
Here, $\{ o_\nu \}$ denotes eigenvalues of $\{ O_\nu \}$ and $P_{\{ o_\nu \}}$ is a projection operator defined by $O_\mu P_{\{ o_\nu \}}= o_\mu P_{\{ o_\nu \}}$, 
$P_{\{ o_\nu \}}P_{\{ o_\nu^\pr \}}=P_{\{ o_\nu \}} \prod_\mu \dl_{o_\mu,o_\mu^\pr}$, and $P_{\{ o_\nu \}} \dg=P_{\{ o_\nu \}}$. 
We suppose $\rho_\tot(0)=\rho(0) \otimes \rho_B(\al_0)$ where 
\bea
\rho_B (\al_t) = \bigotimes_{b} e^{-\beta_b(t) [H_b(t) -\mu_b(t) N_b] }/\Xi_b(\al_t) \la{canon},
\eea 
with $\Xi_b(\al_t)=\tr_b[e^{-\beta_b(t)[ H_b(t) -\mu_b(t) N_b] }]$ and $\tr_b$ denotes the trace of lead $b$. 
Then, $\rho_\tot^\chi(0)=\rho(0) \otimes \sum_{\{ o_\nu \}} P_{\{ o_\nu \}} \rho_B(\al_0) P_{\{ o_\nu \}}$ obeys. 
If all $O_\mu$ are given by $\sum_{b,k,\sig} o^\mu_{bk\sig} c_{bk\sig} \dg c_{bk\sig}$ with real numbers $o^\mu_{bk\sig}$, $P_{\{ o_\nu \}}$ commutes to $\rho_B(0)$ and 
$\rho_\tot^\chi(0)=\rho(0) \otimes \rho_B(\al_0)$ obeys because $\sum_{\{ o_\nu \}} P_{\{ o_\nu \}}=1$.}

Now we move to the \rd{interaction} picture. 
An operator in the \rd{interaction} picture corresponding to $A(t)$ is defined by $A^I(t)=U_0\dg(t)A(t)U_0(t)$ with $\f{dU_0(t)}{dt}=-i[H_S(t)+\sum_{b}H_b(t)]U_0(t)$ and $U_0(0)=1$.
The system reduced density operator in the \rd{interaction} picture is given by $\rho^{I,\chi}(t)=\tr_{\rm{leads}}[\rho_\tot^{I,\chi}(t)]$ where $\rho_\tot^{I,\chi}(t)=U_0(t)\rho_\tot^\chi(t)U_0\dg(t)$ and 
$\tr_{\rm{leads}}$ denotes the trace of lead space. 
\rd{$\rho_\tot^{I,\chi}(t)$ is governed by
\bea
\f{d\rho_\tot^{I,\chi}(t)}{dt} \aeq -i[H_{\rm{int}}^I(t),\rho_\tot^{I,\chi}(t)]_\chi, \la{vN,I}
\eea
with $H_{\rm{int}}=\sum_{b}H_{Sb}$}. Up to the second order perturbation \rd{in $H_{\rm{int}}$}, we obtain
\begin{widetext}
\bea
\rho^{I,\chi}(t+\tau_{\rm{CG}}) 
\aeq \rho^{I,\chi}(t)-\int_t^{t+\tau_{\rm{CG}}} du \int_t^{u} ds \ \tr_{\rm{leads}} \big\{ [ H_{\rm{int}}^I(u), [H_{\rm{int}}^I(s), \rho^{I,\chi}(t) \rho_B (\al_t) ]_\chi ]_\chi \big \} \no\\
\aeqe \rho^{I,\chi}(t) +\tau_{\rm{CG}}\Hat{L}_{\tau_{\rm{CG}}}^\chi(t)\rho^{I,\chi}(t) , \la{FCS-QME_I_CG}
\eea
\end{widetext}
\bl{using the large-reservoir approximation $ \rho_\tot^{I,\chi}(t) \approx \rho^{I,\chi}(t) \otimes \rho_B (\al_t)$ and 
$\tr_{\rm{leads}}[H_{\rm{int}}^I(u)\rho_B (\al_t)]=0$. The arbitrary parameter $\tau_{\rm{CG}}$ $(>0)$ is called the coarse-graining time. }
The coarse-graining approximation\cite{CG,CG13} (CGA) is defined by 
\bea
\f{d}{dt}\rho^{I,\chi}(t)=\Hat{L}_{\tau_{\rm{CG}}}^\chi(t)\rho^{I,\chi}(t). \la{def,CGA}
\eea
If the cycle time of the modulation of control parameters, $\tau$, is much longer than the coarse-graining time \bl{$\tau_{\rm{CG}}$}, 
the superoperator \bl{$\Hat{L}_{\tau_{\rm{CG}}}^\chi(t)$} is described as a function of the set of control parameters at time $t$. 
In this paper, we suppose \bl{$\tau \gg \tau_{\rm{CG}}$}. \bl{Moreover, $\tau_{\rm{CG}}$ should be much shorter than the relaxation time of the system, $\tau_S\sim \f{1}{\Ga}$. 
On the other hand, $\tau_S \ll \tau $ should hold for the adiabatic pump. Hence $\tau_{\rm{CG}} \ll \f{1}{\Ga} \ll \tau$ should hold.}  

In the Schr\"{o}dinger picture, \re{def,CGA} is described as 
\bea
\f{d\rho^\chi(t)}{dt}=-i[H_S(t),\rho^\chi(t)]+\Hat{\Pi}_{\tau_{\rm{CG}}}^\chi(\al_t) \rho^\chi(t) .\la{B7}
\eea
Here, the superoperator \bl{$\Hat{\Pi}_{\tau_{\rm{CG}}}^\chi(\al)$} operates to an operator $\bu$ as 
\begin{widetext}
\bea
\Hat{\Pi}_{\tau_{\rm{CG}}}^\chi(\al) \bu 
\aeq   \sum_b \sum_{\om,\om^\pr }  \sum_{n,m,s,s^\pr} \Big[
 \Phi^-_{b,n s, m s^\pr}(\chi,\tau_{\rm{CG}},\om,\om^\pr) a_{m s^\pr}(\om^\pr) \bu [a_{n s}  (\om)]\dg 
-\half \Phi^-_{b,n s, m s^\pr}(\tau_{\rm{CG}},\om,\om^\pr) \bu [a_{n s}(\om)] \dg a_{m s^\pr}(\om^\pr)\no\\
 \no\\
&&-\half \Phi^-_{b,n s, m s^\pr}(\tau_{\rm{CG}},\om,\om^\pr) [ a_{n s}  (\om)]\dg a_{m s^\pr}(\om^\pr) \bu+\Phi^+_{b,n s, m s^\pr}(\chi,\tau_{\rm{CG}},\om,\om^\pr) [a_{m s^\pr}  (\om^\pr)]\dg \bu a_{n s} (\om) \no\\
&&-\half \Phi^+_{b,n s, m s^\pr}(\tau_{\rm{CG}},\om,\om^\pr) \bu a_{n s}(\om) [a_{m s^\pr}  (\om^\pr)]\dg 
-\half \Phi^+_{b,n s, m s^\pr}(\tau_{\rm{CG}},\om,\om^\pr)  a_{n s}  (\om)[a_{m s^\pr}  (\om^\pr)]\dg \bu  \Big] \no\\ 
 &&\hs{-11mm}-i\sum_b \sum_{\om,\om^\pr }  \sum_{n,m,s,s^\pr} \Big[ -\half \Psi^-_{b,n s, m s^\pr}(\tau_{\rm{CG}},\om,\om^\pr) [ a_{n s}  (\om)]\dg a_{m s^\pr}(\om^\pr) 
+\half \Psi^+_{b,n s, m s^\pr}(\tau_{\rm{CG}},\om,\om^\pr)  a_{n s}  (\om)[a_{m s^\pr}  (\om^\pr)]\dg ,\bu \Big] , \la{CG_go}
\eea
where
\bea
X^\pm(\chi ,\tau_{\rm{CG}},\om,\om^\pr)  \aeq \f{e^{\mp i( \om - \om^\pr)\tau_{\rm{CG}}/2} }{2\pi} \int_{-\infty}^\infty d\Om \hs{0.7mm}  X^\pm(\Om, \chi ) 
\tau_{\rm{CG}}\sinc \big(\f{\tau_{\rm{CG}}(\Om - \om)}{2}\big) \sinc \big(\f{\tau_{\rm{CG}}(\Om - \om^\pr)}{2}\big) ,
\eea
and $X^\pm(\tau_{\rm{CG}},\om,\om^\pr)=X^\pm(\chi=0 ,\tau_{\rm{CG}},\om,\om^\pr)$. Here, $\sinc (x)=\sin x/x$ and $X^\pm(\Om,\chi)$ denotes one of $\Phi_{b,n s,m s^\pr}^\pm(\Om,\chi)$, $\Psi_{b,n s,m s^\pr}^\pm(\Om,\chi) $, where
\bea
\Phi_{b,n s, m s^\pr}^-(\Om,\chi) \aeq 2\pi\sum_{k,\sig}  V_{bk\sig,n s} V_{bk\sig,m s^\pr}^\ast
[1-f_b(\ep_{bk\sig})] e^{i\BL{\chi_{b\sig}} } e^{i\BL{\lm_{b\sig}} \ep_{bk}}  \dl(\ep_{bk\sig}-\Om) ,\la{phi-}\\
\Phi_{b,n s, m s^\pr}^+(\Om,\chi) \aeq 2\pi\sum_{k,\sig} V_{bk\sig,n s}^\ast V_{bk\sig, m s^\pr}
f_b(\ep_{bk\sig}) e^{-i\BL{\chi_{b\sig}} } e^{-i\BL{\lm_{b\sig}} \ep_{bk}}  \dl(\ep_{bk\sig}-\Om),\la{phi+} \\
\Psi_{b,n s, m s^\pr}^-(\Om,\chi) \aeq 2\sum_{k,\sig}  V_{bk\sig,n s} V_{bk\sig,m s^\pr}^\ast  
[1-f_b(\ep_{bk\sig})] e^{i\BL{\chi_{b\sig}} } e^{i\BL{\lm_{b\sig}} \ep_{bk}}  {\rm{P}}\f{1 }{\ep_{bk\sig}-\Om} , \\
\Psi_{b,n s, m s^\pr}^+(\Om,\chi) \aeq 2\sum_{k,\sig} V_{bk\sig,n s}^\ast V_{bk\sig, m s^\pr}
f_b(\ep_{bk\sig}) e^{-i\BL{\chi_{b\sig}} } e^{-i\BL{\lm_{b\sig}} \ep_{bk}}  {\rm{P}}\f{1 }{\ep_{bk\sig}-\Om} ,
\eea
\end{widetext}
with $V_{bk\sig,n s}=\sqrt{\Dl_b}v_{bk\sig,n s}$. $\chi_{b\sig}$ and $\lm_{b\sig}$ denote the counting fields for $N_{b\sig}$ 
and $\sum_k \ep_{bk} c_{bk\sig} \dg c_{bk\sig}$ respectively. 
$f_b(\ep)=[\exp(\be_b(\mu_b-\ep))+1]^{-1}$ is the Fermi distribution function and $\rm{P}$ denotes the Cauchy principal value. The eigenvectors $a_{n s}(\om)$ are given by
\bea
a_{n s}(\om) \aeq \sum_{\al,\be} \dl_{\om_{\be \al},\om} \ke{E_\al}\br{E_\al} a_{n s} \ke{E_\be} \br{E_\be}, \la{Def_a_om}
\eea
with $ \om_{\be\al}= E_\be-E_\al$ and $H_S\ke{E_\al}=E_\al\ke{E_\al}$. $\om$ is one of the elements of 
$\{ \om_{\be\al} \vert \ {  \br{E_\al} a_{n s} \ke{E_\be} \ne 0 \hs{2mm} }^\exists (n,s) \} $.

\rd{We explain the rotating wave approximation (RWA). 
If $H_S$ is time-independent, factors $e^{i(\om-\om^\pr)t}$ appear in \re{BM}. 
The usual RWA \cite{open} is approximating $e^{i(\om-\om^\pr)t}$ as $\dl_{\om,\om^\pr}$ in \re{BM}. 
However, if $H_S$ is time dependent, the generalization of the RWA is unclear. 
In this paper, the RWA is defined as the \rd{limit} $\tau_{\rm{CG}}\to \infty$ of the CGA. 
In this limit [$\tau_{\rm{CG}} \cdot \min_{\om \ne \om^\pr}({\abs{\om-\om^\pr})} \gg 1$], 
$X^\pm(\chi ,\tau_{\rm{CG}},\om,\om^\pr) \approx X^\pm( \om, \chi )\dl_{\om,\om^\pr}$ holds because of the fact that 
$\lim_{\tau_{\rm{CG}} \to \infty}\tau_{\rm{CG}}\sinc \f{\tau_{\rm{CG}}(\Om-\om)}{2} \sinc \f{\tau_{\rm{CG}}(\Om-\om^\pr)}{2}=2\pi \dl_{\om,\om^\pr}\dl(\Om-\om)$. 
If $H_S$ is time independent, this RWA is equivalent to usual RWA.}
 
\rd{In the Born-Markov approximation (without RWA), \re{BM} sometimes violates the non-negativity of $\rho(t)(=\rho^{\chi=0}(t))$ \cite{Mar}. 
The QME of the RWA or the CGA is the Lindblad type (see \re{CG_go} at $\chi=0$) which guarantees the non-negativity \cite{open}. 
In QME of the RWA, the diagonal and off-diagonal elements of  $\rho(t)$ are decoupled. 
Sometimes, the decoupling is justified by the superselection rules\cite{ssr1,ssr2,ssr3}: if two states in the system $\ke{m}$ and $\ke{n}$ differ in an observable 
which is conserved in the total system, unavoidable interactions lead to rapid decay of $\br{m}\rho(t)\ke{n}$. 
The standard example is the electron number which is conserved for \bl{QDs}
coupled to non-superconducting leads. 
The total spin projection is also conserved for unpolarized or collinearly polarized leads when spins do not flip by the interaction $H_{Sb}$. 
Hence, in our setting, Eqs. (\ref{H_b}), (\ref{H_1}) with $v_{bk\sig,ns}\propto \dl_{\sig,s}$, and Eq.(\ref{H_S,1}), which was used after \res{pump,U=0}, 
the decoupling between the diagonal and off-diagonal part of $\br{m_\up m_\dw}\rho(t)\ke{n_\up n_\dw}$ is justified. 
However, in the general system (discussed until \res{SC}), the decoupling can not be justified by the superselection rules. 
In fact, as explained the following, the CGA can break the decoupling. 
However, the QME without decoupling is too hard to analyze analytically.
}

The \rd{FCS-QME} of the CGA (for finite $\tau_{\rm{CG}}$) is more difficult to analyze than that of the RWA. For the one level system \re{H_S,1}, 
the matrix representation of the Liouvillian of the RWA by the number states $\{\ke{n_\up n_\dw} \}$ is block diagonalized
 to the diagonal  part (spanned by  \rd{$\{\ke{n_\up n_\dw}\br{n_\up n_\dw}\}_{n_\up,n_\dw=0,1}$}) and the off-diagonal part
(spanned by \rd{$\{\ke{n_\up n_\dw}\br{m_\up m_\dw} \}_{(n_\up,n_\dw)\ne (m_\up,m_\dw) }$}). 
The diagonal block is given by \re{K_int} and the off-diagonal block is a $(12\times12)$-diagonal matrix.
\bl{However}, one of the CGA has of\bl{f-d}iagonal components which permit transitions between diagonal and off-diagonal if spins can flip by tunneling \rd{$H_{\rm{int}}$}. 
\rd{This is consistent with the above discussion about the superselection rules}.  
Even if spins cannot flip, off-diagonal block is not diagonal. Particularly for $U=0$, the Liouvillian dose not reduce to a summation of one particle 
Liouvillian [$\Hat{K}_\up^\chi(\al)\otimes \Hat{1}_\dw+\Hat{1}_\up \otimes \Hat{K}_\dw^\chi(\al)$]. 
The study of differences between the RWA and the CGA is a future work.

\rd{
Finally, we recognize the Born-Markov approximation (without RWA). From \re{vN,I}, we obtain 
$\rho_\tot^{I,\chi}(t)=\rho_\tot^{I,\chi}(0)-i\int_0^t du \ [H_{\rm{int}}^I(u),\rho_\tot^{I,\chi}(u)]_\chi$. Substituting this to \re{vN,I}, we obtain 
\bea
\f{d\rho_\tot^{I,\chi}(t)}{dt} \aeq
 -i[H_{\rm{int}}^I(t),\rho_\tot^{I,\chi}(0)]_\chi \no\\
&&\hs{-10mm}-[H_{\rm{int}}^I(t),\int_0^t du \ [H_{\rm{int}}^I(u),\rho_\tot^{I,\chi}(u)]_\chi]_\chi. \la{vN,I2}
\eea
Moreover, substituting $\rho_\tot^{I,\chi}(u)=\rho_\tot^{I,\chi}(t)-i\int_t^u ds \ [H_{\rm{int}}^I(s),\rho_\tot^{I,\chi}(s)]_\chi$ to \re{vN,I2} and 
using $ \rho_\tot^{I,\chi}(t) \approx \rho^{I,\chi}(t) \rho_B (\al_t)$ and $\tr_{\rm{leads}}[H_{\rm{int}}^I(t)\rho_B (\al_0)]=0$, 
we obtain
\bea
&&\hs{-10mm}\f{d\rho^{I,\chi}(t)}{dt}= -\int_0^t du \ \tr_{\rm{leads}} \no\\
&& \big\{ [H_{\rm{int}}^I(t),[H_{\rm{int}}^I(u),\rho^{I,\chi}(t) \rho_B (\al_t)]_\chi]_\chi \big\},  \la{RD}
\eea
up to the second order perturbation in $H_{\rm{int}}$. 
Because the integrand decay as $e^{-(t-u)/\tau_B}$ where $\tau_B$ is the relaxation time of the baths, we can replace $\int_0^t du $ to $\int_{-\infty}^t du$, 
and we obtain
\bea
&&\hs{-10mm}\f{d\rho^{I,\chi}(t)}{dt} = -\int_0^\infty ds \ \tr_{\rm{leads}} \no\\
&& \big\{ [H_{\rm{int}}^I(t),[H_{\rm{int}}^I(t-s),\rho^{I,\chi}(t) \rho_B (\al_t)]_\chi]_\chi \big\}. \la{BM}
\eea
This is called the Born-Markov approximation (without RWA). Equations (\ref{RD}) or (\ref{BM}) are also derived from \re{FCS-QME_I_CG}. 
In the  second term of the right side in the first line of \re{FCS-QME_I_CG}, we can replace $\int_t^u ds $ to $\int_0^u ds $ or $\int_{-\infty}^u ds $.
After this replacement, by taking $\f{\partial}{\partial \tau_{\rm{CG}}} \big \vert_{\tau_{\rm{CG}}=0} $, we obtain Eqs. (\ref{RD}) or (\ref{BM}) respectively.
In the Eqs. (\ref{RD}) or (\ref{BM}) (after moving to the Schr\"{o}dinger picture), we can replace 
$\{\al_u\}_{0\le u \le t}$ or $\{\al_{t-s}\}_{0 \le s \le \infty}$ to $\al_t$ because of $\tau_B \ll \tau$.
}

\section{The time evolutions of $c_n^\chi(t)$} \la{co}

In this section, we derive the time evolution equations of $c_n^\chi(t)$ of \re{exp}.
The left hand side of the \rd{FCS-QME}, $\f{d}{dt} \dke{\rho^\chi(t)}=\Hat{K}^\chi(\al_t)  \dke{\rho^\chi(t)}$, is
\bea 
\f{d}{dt} \dke{\rho^\chi(t)} \aeq \sum_n \Big\{ \f{dc_n^\chi(t)}{dt} e^{\Lm_n^\chi (t)}\dke{\rho_n^\chi(\al_t)} \no\\
&&+c_n^\chi(t) e^{\Lm_n^\chi (t)} \lm_n(\al_t) \dke{\rho_n^\chi(\al_t)} \no\\
&&+c_n^\chi(t) e^{\Lm_n^\chi (t)} \f{d}{dt}\dke{\rho_n^\chi(\al_t)}  \Big\} .
\eea
And the right hand side  of the \rd{FCS-QME} is
\bea
\Hat{K}^\chi(\al_t)  \dke{\rho^\chi(t)}
\aeq \sum_n c_n^\chi(t) e^{\Lm_n^\chi (t)}\Hat{K}^\chi(\al_t)  \dke{\rho_n^\chi(\al_t)} \no\\
\aeq \sum_n c_n^\chi(t) e^{\Lm_n^\chi (t)}\lm_n(\al_t)   \dke{\rho_n^\chi(\al_t)}.
\eea
Hence we obtain
\bea
&&\hs{-10mm}\sum_n \Big\{ \f{dc_n^\chi(t)}{dt} e^{\Lm_n^\chi (t)}\dke{\rho_n^\chi(\al_t)} \no\\
&&\hs{5mm}+c_n^\chi(t) e^{\Lm_n^\chi (t)} \f{d\dke{\rho_n^\chi(\al_t)}}{dt}  \Big\} =0. \la{e.c}
\eea
Applying $\dbr{l_m^\chi(\al_t)}$ to \re{e.c}, and using $\dbra l_n^\chi(\al) \dke{\rho_m^\chi(\al)}=\dl_{nm}$, we obtain
\bea
\hs{-7mm}\f{d}{dt}c_m^\chi(t) \aeq -\sum_n c_n^\chi(t) e^{\Lm_n^\chi (t)-\Lm_m^\chi (t)} \dbr{l_m^\chi(\al_t)}\f{d\dke{\rho_n^\chi(\al_t)}}{dt} \la{berry_key}.
\eea

By the way, the time derivative of \re{rig}, $ \Hat{K}^\chi(\al_t) \dke{\rho_n^\chi(\al_t)} =\lm^\chi_n(\al_t) \dke{\rho_n^\chi(\al_t)}$, is 
\bea
&&\hs{-8mm}\f{d \Hat{K}^\chi(\al_t) }{dt} \dke{\rho_n^\chi(\al_t)}+ \Hat{K}^\chi(\al_t) \f{d \dke{\rho_n^\chi(\al_t)} }{dt} \no\\
\aeq \f{ d\lm^\chi_n(\al_t)}{dt} \dke{\rho_n^\chi(\al_t)} +\lm^\chi_n(\al_t) \f{d\dke{\rho_n^\chi(\al_t)} }{dt} .
\eea
Applying $\dbr{l_m^\chi(\al_t)}$ to this equation, we obtain
\bea
&&\hs{-10mm} \dbr{l_m^\chi(\al_t)} \f{d \Hat{K}^\chi(\al_t) }{dt} \dke{\rho_n^\chi(\al_t)}+ \lm_m^\chi(\al_t) \dbr{l_m^\chi(\al_t)} \f{d \dke{\rho_n^\chi(\al_t)} }{dt} \no\\
 \aeq \f{ d\lm^\chi_n(\al_t)}{dt} \dl_{mn}+\lm^\chi_n(\al_t) \dbr{l_m^\chi(\al_t)} \f{d\dke{\rho_n^\chi(\al_t)} }{dt} ,
\eea
and it leads to
\bea
\dbr{l_m^\chi(\al_t)} \f{d\dke{\rho_n^\chi(\al_t)}}{dt} \aeq -\f{\dbr{l_m^\chi(\al_t)} \f{d \Hat{K}^\chi(\al_t) }{dt} \dke{\rho_n^\chi(\al_t)}}{\lm^\chi_m(\al_t)-\lm^\chi_n(\al_t)},
 \la{at_hisyokutai}
\eea
for $\lm^\chi_m(\al_t)\ne \lm^\chi_n(\al_t)$. Substituting this to \re{berry_key}, we obtain
\bea
&&\hs{-5mm}\f{dc^\chi_m(t)}{dt}=-\dbr{l_m^\chi(\al_t)} \f{d}{dt}\dke{\rho_m^\chi(\al_t)}c^\chi_m(t) \no\\
&&\hs{-5mm}+\sum_{n(\ne m)} c^\chi_n(t) e^{\Lm_n^\chi (t)-\Lm_m^\chi (t)} \f{\dbr{l_m^\chi(\al_t)} \f{d \Hat{K}^\chi(\al_t) }{dt} \dke{\rho_n^\chi(\al_t)}}{\lm^\chi_m(\al_t)-\lm^\chi_n(\al_t)} \la{comment2}.
\eea

The above equation can also be written as 
\bea
&&\hs{-5mm}\f{d\tl c^\chi_m(t)}{dt}=\sum_{n(\ne m)} \tl c^\chi_n(t) e^{\Lm_n^\chi (t)-\Lm_m^\chi (t)+\eta_m^\chi(t)-\eta_n^\chi(t)} \no\\
&&\hs{15mm}\times\f{\dbr{l_m^\chi(\al_t)} \f{d \Hat{K}^\chi(\al_t) }{dt} \dke{\rho_n^\chi(\al_t)}}{\lm^\chi_m(\al_t)-\lm^\chi_n(\al_t)} \la{comment3},
\eea
where $\tl c_m^\chi(t)=c_m^\chi(t)e^{\eta_m^\chi(t)}$ with 
\bea
\eta_m^\chi(t)\aeq \int_0^t ds \ \dbr{l_m^\chi(\al_s)} \f{d}{ds}\dke{\rho_m^\chi(\al_s)} \no\\
\aeq \sum_k \int_C d\al^k \  \dbr{l_m^\chi(\al)} \f{\partial}{\partial \al^k}\dke{\rho_m^\chi(\al)}.
\eea
Here, $C$ is the trajectory from $\al_0$ to $\al_t$, $\al^k$ are the $k$th component of the control parameters, and  
$\eta_m^\chi(t)=\mO(1)$ since $\dbr{l_m^\chi(\al_t)} \f{d}{dt}\dke{\rho_m^\chi(\al_t)}=\mO(\om)$ with $\om=2\pi/\tau$. 
In the right hand side of \re{comment3}, the dominant term is $n=0$ if $m\ne 0$ because Re$\lm_0^\chi(\al)>{\rm{Re}}\lm_n^\chi(\al)$. 
Using $\f{d \Hat{K}^\chi(\al_t)} {dt}=\mO(\Ga\om)$, $\lm_n^\chi(\al_t)=\mO(\Ga)$, $e^{\eta_n^\chi(t)}=\mO(1)$ and $c_0^\chi(t)e^{\Lm_0^\chi}=\mO(1)$, we obtain 
\bea
\f{d\tl c^\chi_m(t)}{dt}=\mO( e^{-\Lm_m^\chi (t)}\om ),
\eea 
and 
\bea
c_m^\chi(t)e^{\Lm_m^\chi(t)}\aeq \mO \big(\om \int_0^t ds \ e^{\Lm_m^\chi(t)-\Lm_m^\chi (s)} \big)  \no\\
\aeq \mO\big(\f{\om}{\Ga}\big) \la{ada}.
\eea
For $\chi=0$, \re{ada} is also derived from 
\bea
\rho^a(t) =\rho(t)-\rho_0(\al_t) =\sum_{m\ne 0}c_m(t)e^{\Lm_m(t)}\rho_m(\al_t),
\eea 
and Eqs. (\ref{key6}) and (\ref{order}).

\section{The validity of the adiabatic expansion} \la{val}

In the derivation of the QME with CGA, when going from \re{def,CGA} to \re{B7}, we used the following type of approximation:
\bea
&&\hs{-5mm}\int_t^{t+T}du \int_t^u ds \ G([\al]_s^u;s,u;t) \no\\
\aeqap \int_t^{t+T}du \int_t^u ds \ G([\al_t];s,u;t).
\eea
Here, $G([\al]_s^u,s,u,t) \sim e^{-(u-s)/\tau_B}$ and $[\al]_s^u=(\al_{t^\pr})_{t^\pr=s}^u$ is the control parameters trajectory and $[\al_t]$ is the trajectory which 
$\al_{t^\pr}=\al_t \ (s \le t^\pr \le u)$.
Similarly, in the Bron-Markov approximation (BM), when going from \re{RD} to \re{FCS-QME}, we used
\bea
\int_0^t du \ G([\al]_u^t;u,t;t) \aeqap \int_{-\infty}^t du \ G([\al_t];u,t;t).
\eea
Considering the corrections of the above approximations, the QME are given by
\bea
\f{d\dke{\rho(t)}}{dt} \aeq \mathcal{K}(t)\dke{\rho(t)} ,\la{QM} \\ 
\mathcal{K}(t)\aeq \Hat{K}(\al_t)+\Hat{K}^\pr(t) ,\ \Hat{K}^\pr(t) =\mO(\Ga \om \tau_X) ,
\eea
with $\om=2\pi/\tau$ and $\tau_X=\tau_{\rm{CG}}$ for CGA; $\tau_X=\tau_B$ for BM. The corrections are also discussed in \Ref{a0}.  
The discussions between \re{defR} and \re{order} are correct after replacing ${K}(\al_t) \to \mathcal{K}(t)$, $\mR(\al_t) \to \tl \mR(t)$ and $\rho_0(\al_t) \to \tl \rho_0(t)$. 
Here, $\tl \rho_0(t)$ and $\tl \mR(t)$ are defined by $\mathcal{K}(t)\dke{\tl \rho_0(t)}=0$ and
$\tl \mR(t)\mathcal{K}(t)=1-\dke{\tl \rho_0(t)}\dbr{1}$, respectively. Equation (\ref{key6}) is corrected to
\bea
 \dke{\tl \rho^a(t)}\aeq \tl \mR(t)\f{d\dke{\tl \rho_0(t)}}{dt} +\tl \mR(t)\f{d\dke{\tl \rho^a(t)}}{dt} \no\\
\aeq \sum_{n=1}^\infty \Big[\tl \mR(t)\f{d}{dt} \Big]^n \dke{\tl \rho_0(t)}\e \sum_{n=1}^\infty \dke{\tl \rho^{a(n)}(t)},
\eea
with $\tl \rho^a(t) \defe \rho(t)-\tl \rho_0(t)$. The corrections are given by
\bea
\tl \rho_0=\rho_0[1+\mO(\om \tau_X)] ,\ \tl \mR=\mR[1+\mO(\om \tau_X)] \la{D1},
\eea
and
\bea
\tl \rho^{a(n)}(t)-\rho^{a(n)}(t)=\mO\Big(\big(\f{\om}{\Ga}\big)^n \om \tau_X \Big).
\eea
Next, we consider the reasonable range of $n$ of $\rho^{a(n)}(t)$. Because $\rho^{a(n)}(t)=\mO(\f{\om}{\Ga})^n$ and $\tl \rho_0(t)-\rho_0(\al_t)=\mO(\om \tau_X)$, 
the reasonable range is $n \le n_{\rm{max}}$, where $n_{\rm{max}}$ is determined by
\bea
\big(\f{\om}{\Ga}\big)^{n_{\rm{max}}+1}<\om \tau_X \ll \big(\f{\om}{\Ga}\big)^{n_{\rm{max}}} . \la{n_max}
\eea
Let us consider that reasonable concrete values of the parameters in our model (\res{model}):
$\om=10^p $ MHz, $\Ga=10\ \mu$eV=0.116 K, $1/\Ga=65.8$ ps, $\tau_{\rm{CG}}=1$ ps, and $\tau_B=0.1$ ps. These values lead to
\bea
\hs{-2.5mm}\om \tau_{\rm{CG}}\aeq 10^{-6+p} ,\ \om \tau_B= 10^{-7+p},\ \f{\om}{\Ga}=10^{-4.18+p},
\eea
and $n_{\rm{max}}=[\tl n_{\rm{max}}]$ with
\bea
\tl n_{\rm{max}}=\f{-6+p}{-4.18+p} \ ({\rm{CGA}}),\ \f{-7+p}{-4.18+p} \ (\rm{BM}).
\eea
Here, $[n]$ means the biggest integer below $n$. At $p=0$, $\tl n_{\rm{max}}=1.44$ (CGA), 1.67 (BM) and at $p=3$, $\tl n_{\rm{max}}=2.54$ (CGA), 3.39 (BM).
The larger the nonadiabaticity ($\f{\om}{\Ga}$), the larger $n_{\rm{max}}$ becomes.

\section{Proof of \re{W_R}} \la{proof}

First, using Eqs. (\ref{defR}) and (\ref{touka_sc}), 
we obtain $\dbr{1}W_\mu(\al)\mR(\al)\Hat{K}(\al)=\dbr{1}W_\mu(\al)-\lm_0^\mu(\al)\dbr{1}$. 
Next, $\dbr{l_0(\al)}=\dbr{1}$, $\lm_0(\al)=0$, and Eqs. (\ref{left}) and (\ref{defW}) lead to $\dbr{l_0^\mu(\al)}\Hat{K}(\al)=\lm_0^\mu(\al)\dbr{1}-\dbr{1}W_\mu(\al)$. 
Hence $\big[ \dbr{1}W_\mu(\al)\mR(\al)+\dbr{l_0^\mu(\al)} \big]\Hat{K}(\al)=0 $ and it leads to \re{W_R}. 
To prove \re{W_R} only \re{defR} is required and $\Hat{K}(\al)\mR(\al)=1-\dke{\rho_0(\al)}\dbr{1}$ is not necessary. 
Additionally, the pseudoinverse of the real-time diagrammatic approach \re{RR} satisfies 
\bea
\sum_{\ka}R_{\eta \ka}K_{\ka \zeta}^{(0)} =\dl_{\eta \zeta}-p_\eta^{(0)} \ne \sum_{\ka}K^{(0)} _{\eta \ka}R_{\ka \zeta}, 
\eea
which corresponds to our
\bea
\mR(\al)\Hat{K}(\al)=1-\dke{\rho_0(\al)}\dbr{1}\ne \Hat{K}(\al)\mR(\al). 
\eea

Equation (\ref{W_R}) is shown also as follows. \re{defR} and $\dbr{1}\Hat{K}(\al)=0$ lead to $\Hat{K}(\al)\mR(\al)\Hat{K}(\al)=\Hat{K}(\al)$, which implies
\bea
\Hat{K}(\al)\mR(\al) \aeq  1-\dke{\sig(\al)} \dbr{1}, \ \dbra 1\dke{\sig(\al)}=1 . \la{KR} 
\eea
Applying $\dbr{1}$ to \re{defR}, we obtain $\dbr{1}\mR(\al)\Hat{K}(\al)=0$, which is equivalent to
\bea
\dbr{1} \mR(\al)\aeq \mathcal{C}(\al)\dbr{1}. \la{1_R}
\eea
By the way, differentiating \re{left} for $n=0$ by $i\chi_\mu$, we obtain 
\bea
\dbr{l_0^\mu(\al)}\Hat{K} (\al)+\dbr{1} \Hat{K}^\mu(\al)  \aeq \dbr{1}\lm_0^\mu(\al). 
\eea
Applying $\mR(\al)$ to this equation and using Eqs. (\ref{KR}) and (\ref{1_R}), we obtain\cite{Sagawa, yuge2}
\bea
\dbr{l_0^\mu(\al)}\aeq -\dbr{1} \Hat{K}^\mu(\al)\mR(\al) +c_\mu(\al) \dbr{1} ,\la{Y}\\
c_\mu(\al) \aeq \mathcal{C}(\al)\lm_0^\mu(\al)+\dbr{l_0^\mu(\al)}\sig(\al)\dket \la{Y2}.
\eea
\re{Y} becomes \re{W_R} because of \re{defW}. 
Particularly,  Yuge\cite{yuge2} used
\bea
\mR(\al) \aeq -\lim_{s \to \infty}\int_0^s dt \ e^{\Hat{K}(\al)t}(1-\dke{\rho_{0}(\al)} \dbr{1}) \la{R_Yuge},
\eea
which satisfies Eqs. (\ref{KR}) and (\ref{1_R}) with $\sig(\al)=\rho_0(\al)$, $\mathcal{C}(\al) = 0$ and \re{defR} (in \Ref{yuge2}, $\mathcal{C}(\al)$ was incorrectly set to $-1$).

\end{document}